\documentclass[12pt]{article}
\usepackage{amssymb,amsmath,epsfig}
\allowdisplaybreaks
\begin{document}
\title{\bf Cosmic Expansion and Noether Gauge Symmetries in $f(R,T,R_{\mu\nu}T^{\mu\nu})$ Gravity}

\author{Iqra Nawazish \thanks{iqranawazish07@gmail.com} and M. Sharif \thanks{msharif.math@pu.edu.pk}\\
Department of Mathematics and Statistics, The University of Lahore,\\
Lahore, Pakistan.}

\date{}

\maketitle
\begin{abstract}
The present work explores different evolutionary phases of
isotropically homogeneous and flat cosmos filled with dust fluid in
non-minimally coupled gravity. We consider different models of this
gravity to discuss the presence of symmetry generators together with
conserved integrals using Nother Gauge symmetry scheme. In most of
the cases, we obtain temporal and scaling symmetries that yield
conservation of energy and linear momentum, respectively. In the
absence of contracted Ricci and energy-momentum tensors, we obtain
maximum symmetries but none of them correspond to any standard
symmetry or conservation law. We formulate exact solutions and
construct graphical analysis of standard cosmological parameters. We
observe realistic nature of new models via squared speed of sound,
viability conditions suggested by Dolgov-Kawasaki instability and
state-finder parameters. We investigate the behavior of fractional
densities and check the compatibility with Planck 2018 observational
data. The new models are stable and viable preserving compatibility
with $\Lambda$CDM and Chaplygin gas models. It is concluded that
most of the solutions favor accelerated cosmic expansion.
\end{abstract}

\section{Introduction}

The primal facts and observational evidences reveal that the cosmos
encounters an exponential expansion at the very early stage. This
phase of the universe leads to current accelerated expansion by
following radiation and matter dominated phases. Such variations in
cosmos trigger cosmologists to understand matter distribution and
find possible reasons behind these expanding phases of the universe.
The most compatible explanation is the existence of some exotic
fluid incorporating negative pressure that induces strong
anti-gravitational effects and defines current cosmic expansion. The
origin as well as striking nature of this anti-gravitational force
is yet unknown and consequently, named as dark energy (DE). At
theoretical scales, there are different approaches to deal with
intriguing nature of DE such as modifying Einstein-Hilbert action
that leads to develop modified theories of gravity and dynamical DE
models. The advancements in gravitational part of Einstein-Hilbert
action define higher order minimally as well as non-minimally
coupled gravitational theories, i.e., $f(R)$, $f(R,T)$ and
$f(R,T,R_{\mu\nu}T^{\mu\nu})$ ($R_{\mu\nu}$, $T^{\mu\nu}$, $R$ and
$T$, represent the Ricci and energy-momentum tensors with their
traces, respectively) theories. The impact of minimal coupling
between curvature and matter variables yields intriguing
phenomenologies and solution to many cosmological problems \cite{2}.

The revolutionary idea of introducing non-minimal interactions
between geometric and matter contents put forward fascinating
approaches to investigate different cosmological scenarios and
current state of cosmos. Bertolami et al. \cite{3} proposed a
generic function admitting non-minimal interactions between scalar
curvature and matter Lagrangian. Harko et al. \cite{4} extended this
non-minimal coupling by replacing matter Lagrangian with trace of
the energy-momentum tensor, referred to as $f(R,T)$ theory. Such
advancements not only deal with cosmic evolution and current
expansion but also suggest efficient approaches to study dark matter
in galaxies, natural conditions for early universe and existence of
feasible cosmological configurations \cite{5}. Sharif and Zubair
\cite{6} followed this idea of non-minimal coupling to interpret
thermodynamical picture, stability criteria, reconstruction of some
new DE models and exact solutions of isotropic/anisotropic
cosmological models.

In the presence of electromagnetic field, the non-minimal
interactions between geometric and matter variables identically
disappear from the field equations of $f(R,T)$ gravity for $T=0$ and
recovers $f(R)$ gravity. This aspect motivates to introduce a
generalized version of $f(R,T)$ theory by taking into account
non-minimal coupling between contracted Ricci and energy-momentum
tensors known as $f(R,T,R_{\mu\nu}T^{\mu\nu})$ gravitational theory
\cite{b1}. Unlike $f(R,T)$ theory, the extended version preserves
coupling of electromagnetic field with contracted Ricci and
energy-momentum tensors for $T=0$ \cite{m1}. In non-minimally
coupled theories, the energy-momentum tensor remains non-conserved
due to the presence of an extra force that deviates massive test
particles and also introduce instabilities against local
perturbations. Using Dolgov-Kawasaki instability criteria, some
constraints are introduced whose viability eliminates these
instabilities \cite{b1a}. Sharif and Zubair \cite{7} discussed
thermodynamical laws and viability of energy constraints for
different cosmological models. Different cosmological aspects like
gravitational collapse, dynamical instability, cosmic evolution and
exact solutions for self-gravitating objects are significantly
studied in the frame-work of $f(R,T,R_{\mu\nu}T^{\mu\nu})$ gravity
\cite{ab}.

In minimally/non-minimally coupled gravitational theories, the exact
solutions to the non-linear field equations come up with significant
approaches to understand cosmic evolution, configurations and matter
contributions. Sebastiani and Zerbini \cite{a1} formulated
non-trivial exact solutions for static spherically symmetric
structure in $f(R)$ gravity. In the presence of scalar field,
Maharaj et al. \cite{a12} found solutions corresponding to de
Sitter, oscillating, accelerating, decelerating and contracting
cosmos in the same gravity. Sharif and Zubair \cite{11} determined
solutions for power-law and exponential anisotropic cosmological
models in $f(R,T)$ theory. Harko and Lake \cite{8} calculated exact
solutions to discuss the effect of non-minimal coupling between
scalar curvature and matter Lagrangian on cylindrical model. Shamir
and Raza \cite{9} obtained two solutions characterizing cosmic
string and non-null electromagnetic field in the background of
$f(R,T)$ gravity. Shamir \cite{12} found three unique solutions for
Bianchi I universe model and discussed their physical behavior via
cosmological parameters in the same gravity.

The technique of Noether symmetry puts forward an interesting way to
identify symmetries and relative conserved entities of cosmological
systems. This approach significantly reduces the complexity of
non-linear higher order partial differential equations and yield
corresponding exact solutions. The existence of symmetries and
conservation laws enhance physical worth of modified theories as if
a theory does not incorporate any symmetry or conserved quantity
then it may refer to as nonphysical theory. Capozziello et al.
\cite{20} considered constant Ricci scalar and power-law $f(R)$
model to find exact solutions for static spherically symmetric
metric using Noether symmetry technique. Hussain et al. \cite{23}
evaluated Noether point symmetries of flat FRW metric with same
$f(R)$ model and zero boundary term. Shamir et al. \cite{24}
extended their work for non-zero boundary term and obtained some
extra symmetries.

Atazadeh and Darabi \cite{15} followed this technique to reconstruct
viable $f(\mathrm{T})$ models ($\mathrm{T}$ represents torsion) that
corresponds to power-law expansion. Momeni et al. \cite{a3} solved
over determining system of mimetic $f(R)$ gravity to get Noether
point symmetry together with conserved charge whereas they found
power-law solution that explains decelerating cosmic expansion in
$f(R,T)$ theory. Sharif and Fatima \cite{a2} established symmetry
generators and conserved entities for both vacuum as well as
non-vacuum flat isotropic homogeneous cosmological model in
$f(\mathcal{G})$ gravity ($\mathcal{G}$, referred to Gauss-Bonnet
invariant). We have discussed the existence of Noether symmetries
with conserved quantities and evaluated some exact solutions of
anisotropic universe model in the background of $f(R)$ and $f(R,T)$
theories \cite{17}. Besides cosmological evolution and current
expansion, we have also studied cosmological configurations like
wormhole whose stability as well as viability is examined via
constant and variable red-shift functions in both theories
\cite{16}. Sharif et al. \cite{18} established some realistic and
viable wormholes for both exponential as well as quadratic
$f(\mathcal{G})$ models.

Motivated from above significant outcomes and growing interest in
cosmological aspects, it would be interesting to study cosmic
expansion and evolution in the background of non-minimally coupled
$f(R,T,R_{\mu\nu}T^{\mu\nu})$ theory. In this paper, we consider
flat isotropic dust cosmological model to evaluate Noether point
symmetries, conserved integrals and some solutions using Noether
Gauge symmetry scheme. We construct graphical analysis of standard
cosmological parameters and fractional densities. To study model
feasibility, we also investigate the behavior of squared speed of
sound, viability constraints and state-finder parameters. The format
of the paper is given as follows. In section \textbf{2}, we discuss
basic formulation of $f(R,T,R_{\mu\nu}T^{\mu\nu})$ gravity and
Noether gauge symmetry scheme. In sections \textbf{3-6}, we obtain
Noether gauge symmetries, conserved quantities and exact solutions.
Furthermore, we establish cosmological analysis via standard
cosmological parameters and corresponding graphical illustration. In
the last section, we provide a summary of our results.

\section{$f(R,T,R_{\mu\nu}T^{\mu\nu})$ Gravity and Noether Gauge Symmetry Approach}

For non-minimally interacting curvature and matter variables, the
action of this gravity is defined as \cite{b1}
\begin{equation}\label{1}
\mathcal{I}=\frac{1}{2\kappa^2}\int
d^4x\sqrt{-g}[f(R,T,R_{\mu\nu}T^{\mu\nu})+ \mathcal{L}_m],
\end{equation}
where $f$ represents a generic function inducing non-minimal
coupling between scalar curvature, matter and contracted Ricci and
energy-momentum tensors while $\kappa$, $g$ and $\mathcal{L}_m$
describe coupling constant, determinant of the metric tensor
($g_{\mu\nu}$) and Lagrangian relative to ordinary matter,
respectively. For the sake of simplicity, we consider
$R_{\mu\nu}T^{\mu\nu}=Q$. The matter Lagrangian depending on the
metric tensor leads to the following form of the energy-momentum
tensor
\begin{equation}\label{2}
T_{\mu\nu}=-\frac{2}{\sqrt{-g}}\frac{\delta(\sqrt{-g}\mathcal{L}_m)}{\delta
g^{\mu\nu}}=g_{\mu\nu}\mathcal{L}_{m}
-\frac{2\delta\mathcal{L}_{m}}{\delta g^{\mu\nu}}.
\end{equation}
For $\kappa^2=1$, the variation of action (\ref{1}) with respect to
the metric tensor yields non-linear partial differential field
equation as follows
\begin{eqnarray}\nonumber
&&(f_{_R}-f_{_Q}\mathcal{L}_{m})G_{\mu\nu} + \left[\Box
f_{_R}+\frac{1}{2}Rf_{_R}-\frac{1}{2}f+f_{_T}\mathcal{L}_{m}+
\frac{1}{2}\nabla_{\alpha}\nabla_{\beta}\left(f_{_Q}T^{\alpha\beta}
\right)\right]g_{\mu\nu}\\\nonumber
&&-\nabla_{\mu}\nabla_{\nu}f_{_R}
+\frac{1}{2}\Box(f_{_Q}T_{\mu\nu})+2f_{_Q}R_{\alpha(\mu}T_{\nu)}^{\alpha}
-\nabla_{\alpha}\nabla_{(\mu}[T_{\nu)}^{\alpha}f_{_Q}] -2
(f_{_T}g^{\alpha\beta}\\\label{3}&&+f_{_Q}R^{\alpha\beta}
)\frac{\partial^{2}\mathcal{L}_{m}}{\partial g^{\mu\nu}\partial
g^{\alpha\beta}} =(1+f_{_T}+\frac{1}{2}Rf_{_Q})T_{\mu\nu}.
\end{eqnarray}
Here $G_{\mu\nu}$ identifies as the Einstein tensor, $\nabla_{\mu}$
defines covariant derivative, $\Box=\nabla_{\mu}\nabla^{\mu}$
whereas $f_{_R},~f_{_T}$ and $f_{_Q}$ denote derivative of the
generic function corresponding to $R,~T$ and $Q$, respectively.

The equivalent form of the above field equations to the Einstein
field equations is given by
\begin{equation}\label{4}
G_{\mu\nu}= R_{\mu\nu}-\frac{1}{2}Rg_{\mu\nu}=T_{\mu\nu}^{eff},
\end{equation}
where the effective energy-momentum tensor $T_{\mu\nu}^{eff}$
incorporates matter and higher order curvature terms given as
\begin{eqnarray}\nonumber
T_{\mu\nu}^{eff}&=&\frac{1}{f_{_R}-f_{_Q}\mathcal{L}_{m}}[(1+f_{_T}+\frac{1}{2}
Rf_{_Q})T_{\mu\nu}+\{\frac{1}{2}(f-Rf_{_R})-\mathcal{L}_{m}f_{_T}\\\nonumber&-&
\frac{1}{2}\nabla_{\alpha}\nabla_{\beta}(f_{_Q}T^{\alpha\beta})\}g_{\mu\nu}
-(g_{\mu\nu}\Box-\nabla_{\mu}\nabla_{\nu})f_{_R}-\frac{1}{2}\Box(f_{_Q}T_{\mu\nu})
\\\label{5}&+&\nabla_{\alpha}\nabla_{(\mu}[T_{\nu)}^{\alpha}f_{_Q}]-2f_{_Q}
R_{\alpha(\mu}T_{\nu)}^{\alpha}+2(f_{_T}g^{\alpha\beta}+f_{_Q}R^{\alpha\beta})
\frac{\partial^{2}\mathcal{L}_{m}}{\partial g^{\mu\nu}\partial
g^{\alpha\beta}}].
\end{eqnarray}
The contraction of Eq.(\ref{3}) relative to the metric tensor leads
to construct a correspondence between trace of geometric and matter
parts as follows
\begin{eqnarray*}\nonumber
&&(f_{_R}+f_{_Q}\mathcal{L}_m)R+\nabla_{\alpha}\nabla_{\beta}(f_{_Q}T^{\alpha\beta})
+4f_{_T}\mathcal{L}_m-2f+3\Box
f_{_R}+\frac{1}{2}\Box(f_{_Q}T)\\\nonumber&&+2f_{_Q}R_{\alpha\beta}T^{\alpha\beta}-2g^{\mu\nu}
(f_{_T}g^{\alpha\beta}+f_{_Q}R^{\alpha\beta})\frac{\partial^{2}\mathcal{L}_{m}}{\partial
g^{\mu\nu}\partial g^{\alpha\beta}}=(1+f_{_T}+\frac{1}{2}Rf_{_Q})T.
\end{eqnarray*}
In the presence of non-minimal curvature-matter interactions, the
covariant derivative of Eq.(\ref{3}) fails to satisfy conservation
of the energy-momentum tensor yielding
\begin{eqnarray*}\label{T}
&&\nabla^{\mu}T_{\mu\nu}=\frac{1}{1+f_{_T}+(Rf_{_Q})/2}\left[\nabla_{\mu}
(f_{_Q}T_{\sigma\nu}R^{\sigma\mu})-\frac{1}{2}(f_{_Q}f_{_T}R_{\alpha\beta}
g_{\alpha\beta})\nabla_{\nu}T^{\alpha\beta}\right.\\\nonumber&&\left.
+\nabla_{\nu}(\mathcal{L}_mf_{_T})
-G_{\mu\nu}\nabla^{\mu}(f_{_Q}\mathcal{L}_m)-\frac{1}{2}(\nabla^{\mu}(Rf_{_Q})
+2\nabla^{\mu}f_{_T})T_{\mu\nu}\right].
\end{eqnarray*}

The non-conserved energy-momentum tensor introduces an extra force
given by
\begin{eqnarray*}\label{T}
&&\frac{d^2x^{\alpha}}{ds^2}+\Gamma^{\alpha}_{\mu\nu}u^{\mu}u^{\nu}=
\mathcal{F}^{\alpha},
\end{eqnarray*}
where ${F}^{\alpha}$ defines extra force orthogonal to four velocity
of the massive particles. In $f(R,T,Q)$ theory, it becomes
\begin{eqnarray*}\label{T}
&&\mathcal{F}^{\alpha}=\frac{h^{\alpha\beta}}
{(\rho_m+p_m)(1+2f_T+Rf_{_{RT}})}[(f_T+Rf_{_{RT}})
\nabla_{\beta}\rho_m-(1+3f_T)\nabla_{\beta}p_m\\\nonumber&&-(\rho_m+p_m)f_{_{RT}}
R^{\gamma\delta}(\nabla\beta h_{\gamma\delta}-2\nabla\delta
h_{\gamma\beta})-f_{_{RT}}R_{\gamma\delta}h^{\gamma\delta}
\nabla_{\beta}(\rho_m+p_m)].
\end{eqnarray*}
The minimally coupled gravitational theories are strongly supported
by the equivalence principle as it passes solar system tests in weak
gravitational fields whereas non-minimally coupled theories
incorporate an additional force that explicitly violates this
principle. Recent observations of Abell Cluster A586 claim that
violation of the equivalence principle is not the only criteria to
rule out any gravitational theory as the equivalence principle test
is constrained under the influence of weak gravitational force and
results may vary in the presence of strong gravitational fields or
interactions of DE or DM \cite{m2}. In modified theories of gravity,
the matter Lagrangian plays a significant role to study
conserved/non-conserved nature of matter. Different choices of
matter Lagrangian interacting with curvature invariant lead to
investigate the impact of geodesic as well as non-geodesic motion
that unravel different cosmological issues and also elaborate
physical constraints of a theory. In $f(R,\mathcal{L}_m)$ theory,
the additional force vanishes for $\mathcal{L}_m=p_m$ \cite{b2}. In
$f(R,T)$ theory, the geodesic lines of motion are followed for
perfect fluid distribution with $f_{T}(R,T)=0$ whereas the effect of
extra force can be neglected in the presence of dust particles even
with $f_{T}(R,T)\neq0$. In $f(R,T,Q)$ theory, the additional force
cannot be avoided even for dust particles or any particular choice
of matter Lagrangian due to explicit dependence on the Ricci tensor.
This dependence helps to explore evolutionary phases of cosmic
regimes with strong curvature. The effect of non-geodesic particles
can be ignored only for non-interacting curvature and matter
variables, i.e., $f_{T}(R,T)=f_{_{RT}}=0$ \cite{4}.

The ordinary matter of cosmos is assumed to be distributed with
perfect fluid whose energy-momentum takes the form
\begin{equation*}\label{F}
T_{\mu\nu}=(\rho_m+p_m)v_\mu v_\nu+p_mg_{\mu\nu},\quad
v_\mu=(-1,0,0,0),
\end{equation*}
where $\rho_m$ and $p_m$ refer to energy density and pressure,
respectively whereas $v_\mu$ describes four velocity of the ordinary
matter. The non-conserved nature of matter is independent of matter
Lagrangian as the extra force does not disappear even for
$\mathcal{L}_m=p_m$ or $\mathcal{L}_m=-\rho_m$. Therefore, the
selection of matter Lagrangian is not unique and we consider
$\mathcal{L}_m=-\rho_m$ for perfect fluid distribution.

For isotropic and homogenous flat cosmos, the cosmological model is
described as
\begin{equation}\label{6a}
ds^2=-dt^2+a^2(t)(dx^2+dy^2+dz^2),
\end{equation}
where the scale factor $a(t)$ represents cosmic expansion along
$x,~y$ and $z$-directions. In order to construct point-like
Lagrangian for the action (\ref{1}), we use Lagrange multiplier
approach leading to the following form
\begin{eqnarray}\nonumber
\mathcal{I}&=&\int\sqrt{-g}[f(R,T,Q)-\zeta(R-\bar{R})-\eta(T-\bar{T})
-\vartheta(Q-\bar{Q})+\rho_m(a)]dt,\\\label{6}
\bar{R}&=&6\left(\frac{\ddot{a}}{a}
+\frac{\dot{a}^2}{a^2}\right),\quad\bar{T}=3p_m-\rho_m,\quad
Q=-\frac{3\ddot{a}\rho_m}{a}+\frac{3p_m}{a^2}\left(2\dot{a}^2+a\ddot{a}\right).
\end{eqnarray}
Here $\sqrt{-g}=a^3$, $\zeta,~\eta$ and $\vartheta$ are scalar
curvature terms whereas $\bar{R},~\bar{T}$ and $\bar{Q}$ refer to
dynamical constraint. Varying the above action relative to $R,~T$
and $Q$, we obtain $\zeta=f_{_R},~\eta=f_{_T}$ and
$\vartheta=f_{_Q}$ while integrating the second order derivatives in
Eq.(\ref{6}) leads to the following Lagrangian
\begin{eqnarray}\nonumber
&&\mathcal{L}(a,R,T,Q,\dot{a},\dot{R},\dot{T},\dot{Q})=a^3[f(R,T,Q)-Rf_{_R}
-Tf_{_T}-Qf_{_Q}-(\rho_m-3p_m)\\\nonumber&&\times
f_{_T}-\rho_m]-6a\dot{a}^2f_{_R}
-6a^2\dot{a}\dot{R}f_{_{RR}}-6a^2\dot{a}\dot{T}f_{_{RT}}
-6a^2\dot{a}\dot{Q}f_{_{RQ}}+3a\dot{a}^2f_{_Q}(2\rho_m\\\label{7}&&-ap_m'
+a\rho_m')+3a^2\dot{a}(\rho_m-p_m)[\dot{R}f_{_{RQ}}+\dot{T}f_{_{TQ}}
+\dot{Q}f_{_{QQ}}],
\end{eqnarray}
In a dynamical system, Hamiltonian equation interprets the total
energy of a system while equation of motion is measured through
Lagrange equation. Mathematically, these equations are defined as
\begin{eqnarray}\nonumber
\mathcal{H}=\Sigma_i\dot{q}^ip_i-\mathcal{L},\quad
\frac{\partial\mathcal{L}}{\partial
q^i}-\frac{d}{dt}\left(\frac{\partial\mathcal{L}}{\partial
\dot{q}^i}\right)=0 ,
\end{eqnarray}
where $q^i,~\dot{q}^i$ and $p^i$ describe generalized co-ordinates,
velocity and momentum, respectively. For Eq.(\ref{7}), the
Hamiltonian equation is
\begin{eqnarray}\nonumber
\mathcal{H}&=&-\frac{3\dot{a}^2}{a^2}+\frac{1}{2(f_{_R}-\rho_mf_{_Q})}\left[f(R,T,Q)
-Rf_{_R}-Tf_{_T}-f_T(\rho_m-3p_m)\right.\\\nonumber&+&\rho_m-Qf_{_Q}+\left.3\dot{a}H
(\rho_m'-p_m')f_{_Q}+3H(\rho_m-p_m)\{\dot{R}f_{_{RQ}}
+\dot{T}f_{_{TQ}}+\dot{Q}f_{_{QQ}}\}\right.\\\label{8}&-&\left.6H\{\dot{R}f_{_{RR}}
+\dot{T}f_{_{RT}}+\dot{Q}f_{_{RQ}}\}\right].
\end{eqnarray}
The Hamiltonian equation is also used to evaluate total energy
density of the dynamical system for constraint $\mathcal{H}=0$. For
generalized co-ordinates $q_i=\{a,R,T,Q\}$ and corresponding
Lagrangian (\ref{7}), the equations of motion turn out to be
\begin{eqnarray}\nonumber
&&\dot{a^2}+2a\ddot{a}+\frac{a^2}{2(f_{_R}-\rho_mf_{_Q})}\left[f(R,T,Q)
-Rf_{_R}-Tf_{_T}-Qf_{_Q}-f_T(\rho_m-3p_m)\right.\\\nonumber&&-\left.\rho_m-
\frac{a}{3}(f_T(\rho_m'-3p_m')+\rho_m')+2H\{2\dot{f}_{_R}
-\dot{f}_{_Q}(2\rho_m+a\rho_m'-ap_m')\}\right.\\\nonumber&&+\left.2(\ddot{R}f_{_{RR}}+
\ddot{T}f_{_{RT}}+\ddot{Q}f_{_{RQ}}+\dot{R}\dot{f}_{_{RR}}+
+\dot{T}\dot{f}_{_{RT}}+\dot{Q}\dot{f}_{_{RQ}})+(\ddot{a}+2aH^2)
\right.\\\nonumber&&\times\left.
(\rho_m'-p_m')f_{_Q}-aH^2f_{_Q}(6\rho_m'-2p_m'+a\rho_m''-ap_m'')
+(\rho_m-p_m)\right.\\\label{9}&&\times\left.(\ddot{R}f_{_{RQ}}+
\ddot{T}f_{_{TQ}}+\ddot{Q}f_{_{QQ}}+\dot{R}\dot{f}_{_{RQ}}
+\dot{T}\dot{f}_{_{TQ}}+\dot{Q}\dot{f}_{_{QQ}})\right]=0,
\\\nonumber&&-a^3(Rf_{_{RR}}+Tf_{_{RT}}+Qf_{_{RQ}}+f_{_{RT}}(\rho_m-3p_m))
+6a\dot{a}^2(f_{_{RR}}+\rho_mf_{_{RQ}})\\\label{10}&&
+6a^2\ddot{a}f_{_{RR}}-3af_{_{RQ}}(\rho_m-p_m)(2\dot{a}^2+a\ddot{a})=0,
\\\nonumber&&-a^3(Rf_{_{RT}}+Tf_{_{TT}}+Qf_{_{TQ}}+f_{_{TT}}(\rho_m-3p_m))
+6a\dot{a}^2(f_{_{RT}}+\rho_mf_{_{TQ}})\\\label{11}&&
+6a^2\ddot{a}f_{_{RT}}-3af_{_{TQ}}(\rho_m-p_m)(2\dot{a}^2+a\ddot{a})=0,
\\\nonumber&&-a^3(Rf_{_{RQ}}+Tf_{_{TQ}}+Qf_{_{QQ}}+f_{_{TQ}}(\rho_m-3p_m))
+6a\dot{a}^2(f_{_{RQ}}+\rho_mf_{_{QQ}})\\\label{12}&&
+6a^2\ddot{a}f_{_{RQ}}-3af_{_{QQ}}(\rho_m-p_m)(2\dot{a}^2+a\ddot{a})=0.
\end{eqnarray}

In order to formulate exact solutions of the above non-linear
partial differential equations, we use Noether symmetry approach
that not only reduces the complexity of the equations but also
helpful to understand enigmatic behavior of DE \cite{b4}. This
technique is based on a well-known Noether theorem which connects
symmetries induced by symmetry generators and conservation laws
under the invariance of Lagrangian. The existence of symmetries and
corresponding conserved entities also support the physical
interpretation of minimal as well as non-minimal gravitational
theories. For the generalized co-ordinates $q^i$ and affine
parameter $t$, the symmetry generator, corresponding invariance
condition and Noether first integral are defined as
\begin{eqnarray}\nonumber
Y&=&\chi(t,q^i)\partial_t+\xi^j(t,q^i)\partial_{q^j},\quad
Y^{[1]}\mathcal{L}+(D\chi)\mathcal{L}=D\mathcal{B}(t,q^i),\\\label{13}
I&=&\mathcal{B}-\chi\mathcal{L}-(\xi^j-\dot{q}^j\chi)
\frac{\partial\mathcal{L}}{\partial\dot{q}^j},
\end{eqnarray}
where $\mathcal{B}$ identifies boundary term that ensures the
presence of some extra symmetries also referred as Noether Gauge
symmetry whereas the first order prolongation $Y^{[1]}$ and total
derivative $D$ relative to symmetry generator $Y$ are given by
\begin{eqnarray}\label{15}
Y^{[1]}=Y+(\chi^j,_t+\chi^j,_i\dot{q}^i-\xi,_t\dot{q}^j
-\xi,_i\dot{q}^i\dot{q}^j)\frac{\partial}{\partial\dot{q}^j},\quad
D=\partial_t+\dot{q}^i\partial_{q^i}.
\end{eqnarray}

We define the vector field corresponding to affine parameter and
generalized co-ordinates relative to the Lagrangian (\ref{7}) as
follows
\begin{eqnarray}\label{18}
Y&=&\chi\partial_{t}+\alpha\partial_{a}+\beta\partial_{R}
+\phi\partial_{T}+\psi\partial_{Q}.
\end{eqnarray}
Here $\chi,~\alpha,~\beta,~\phi$ and $\psi$ are unknown functions
with respect to configuration space $\mathcal{Q}=\{t,~a,~R,~T,~Q\}$.
For the tangent space
$\mathcal{T}=\{t,~a,~\dot{a},~R,~\dot{R},~T,~\dot{T},\\Q,~\dot{Q}\}$,
the first order prolongation takes the form
\begin{equation}\label{19}
Y^{[1]}=\chi\partial_{t}+\alpha\partial_{a}+\beta\partial_{R}
+\phi\partial_{T}+\psi\partial_{Q}+\dot{\alpha}\partial_{\dot{a}}
+\dot{\beta}\partial_{\dot{R}}+\dot{\phi}\partial_{\dot{T}}
+\dot{\psi}\partial_{\dot{Q}}.
\end{equation}
Using Eq.(\ref{15}), the time derivative of the above unknown
functions yield
\begin{eqnarray}\nonumber
\dot{\alpha}&=&\alpha_{_t}+\dot{a}\alpha_{_a}
+\dot{R}\alpha_{_R}+\dot{T}\alpha_{_T}
+\dot{Q}\alpha_{_Q}-\dot{a}\left\{\chi_{_t}+\dot{a}\chi_{_a}
+\dot{R}\chi_{_R}
+\dot{T}\chi_{_T}+\dot{Q}\chi_{_Q}\right\},\\
\nonumber \dot{\beta}&=&\beta_{_t}+\dot{a}\beta_{_a}
+\dot{R}\beta_{_R}+\dot{T}\beta_{_T}
+\beta\alpha_{_Q}-\dot{R}\left\{\chi_{_t}+\dot{a}\chi_{_a}
+\dot{R}\chi_{_R}
+\dot{T}\chi_{_T}+\dot{Q}\chi_{_Q}\right\},\\
\nonumber \dot{\phi}&=&\phi_{_t}+\dot{a}\phi_{_a}
+\dot{R}\phi_{_R}+\dot{T}\phi_{_T}
+\dot{Q}\phi_{_Q}-\dot{T}\left\{\chi_{_t}+\dot{a}\chi_{_a}
+\dot{R}\chi_{_R}
+\dot{T}\chi_{_T}+\dot{Q}\chi_{_Q}\right\},\\
\nonumber \dot{\psi}&=&\psi_{_t}+\dot{a}\psi_{_a}
+\dot{R}\psi_{_R}+\dot{T}\psi_{_T}
+\dot{Q}\psi_{_Q}-\dot{Q}\left\{\chi_{_t}+\dot{a}\chi_{_a}
+\dot{R}\chi_{_R} +\dot{T}\chi_{_T}+\dot{Q}\chi_{_Q}\right\}.
\end{eqnarray}
To construct non-linear system of over-determined equations, we
insert Eq.(\ref{7}), (\ref{18}) and (\ref{19}) with time derivative
of unknown functions in invariance condition (\ref{13}). After
comparing coefficients of different products of generalized
co-ordinates and their derivatives, we obtain
\begin{eqnarray}\label{20}
&&\alpha_{_{R}}[2f_{_{RR}}-(\rho_m-p_m)f_{_{RQ}}]=0,
\\\label{21a}&&\alpha_{_{T}}[2f_{_{RT}}-(\rho_m-p_m)f_{_{TQ}}]=0,
\\\label{21}&&\alpha_{_{Q}}[2f_{_{RQ}}-(\rho_m-p_m)f_{_{QQ}}]=0,
\\\label{26}&&3a^2\alpha,_{_t}[2f_{_{RR}}-f_{_{RQ}}(\rho_m'-p_m')]=-\mathcal{B}_{_R},
\\\label{27}&&3a^2\alpha,_{_t}[2f_{_{RT}}-f_{_{TQ}}(\rho_m'-p_m')]=-\mathcal{B}_{_T},
\\\label{28}&&3a^2\alpha,_{_t}[2f_{_{RQ}}-f_{_{QQ}}(\rho_m'-p_m')]=-\mathcal{B}_{_Q},
\\\label{22}&&\chi_{_{a}}[2f_{_{R}}-2\rho_mf_{_{Q}}-af_{_{Q}}(\rho_m'-p_m')]=0,
\\\label{23}&&\chi_{_{R}}[2f_{_{RR}}-f_{_{RQ}}-2f_{_{RQ}}(\rho_m'-p_m')]=0,
\\\label{24}&&\chi_{_{T}}[2f_{_{RT}}-f_{_{TQ}}-2f_{_{TQ}}(\rho_m'-p_m')]=0,
\\\label{25}&&\chi_{_{Q}}[2f_{_{RQ}}-f_{_{QQ}}-2f_{_{QQ}}(\rho_m'-p_m')]=0,
\\\label{29}&&\alpha_{_{R}}[2f_{_{RT}}-(\rho_m-p_m)f_{_{TQ}}]
+\alpha_{_{T}}[2f_{_{RR}}-(\rho_m-p_m)f_{_{RQ}}]=0,
\\\label{30}&&\alpha_{_{R}}[2f_{_{RQ}}-(\rho_m-p_m)f_{_{QQ}}]
+\alpha_{_{Q}}[2f_{_{RR}}-(\rho_m-p_m)f_{_{RQ}}]=0,
\\\label{31}&&\alpha_{_{Q}}[2f_{_{RT}}-(\rho_m-p_m)f_{_{TQ}}]
+\alpha_{_{T}}[2f_{_{RQ}}-(\rho_m-p_m)f_{_{QQ}}]=0,
\\\nonumber&&\chi_{_{T}}[2f_{_{RR}}-f_{_{RQ}}-2f_{_{RQ}}(\rho_m'-p_m')]
+\chi_{_{R}}[2f_{_{RT}}-f_{_{TQ}}-2f_{_{TQ}}(\rho_m'-p_m')]=0,\\\label{32}
\\\nonumber&&\chi_{_{R}}[2f_{_{RQ}}-f_{_{QQ}}-2f_{_{QQ}}(\rho_m'-p_m')]
+\chi_{_{Q}}[2f_{_{RR}}-f_{_{RQ}}-2f_{_{RQ}}(\rho_m'-p_m')]=0,\\\label{33}
\\\nonumber&&\chi_{_{T}}[2f_{_{RQ}}-f_{_{QQ}}-2f_{_{QQ}}(\rho_m'-p_m')]
+\chi_{_{Q}}[2f_{_{RT}}-f_{_{TQ}}-2f_{_{TQ}}(\rho_m'-p_m')]=0,\\\label{34}
\\\nonumber&&\chi_{_{R}}[2f_{_{R}}-2\rho_mf_{_{Q}}-af_{_{Q}}(\rho_m'-p_m')]
+a\chi_{_{a}}[2f_{_{RR}}-f_{_{RQ}}-2f_{_{RQ}}(\rho_m-p_m)]=0,
\\\label{35}\\\nonumber&&\chi_{_{T}}[2f_{_{R}}-2\rho_mf_{_{Q}}-af_{_{Q}}
(\rho_m'-p_m')]+a\chi_{_{a}}[2f_{_{RT}}-f_{_{TQ}}-2f_{_{TQ}}(\rho_m-p_m)]=0,
\\\label{36}\\\nonumber&&\chi_{_{Q}}[2f_{_{R}}-2\rho_mf_{_{Q}}-af_{_{Q}}
(\rho_m'-p_m')]+a\chi_{_{a}}[2f_{_{RQ}}-f_{_{QQ}}-2f_{_{QQ}}(\rho_m-p_m)]=0,
\\\label{37}\\\nonumber&&6a\alpha_{_{t}}[2f_{_{R}}+f_{_{Q}}(2\rho_m
+a\rho_m'-ap_m')]-3a^2\beta_{_{t}}[2f_{_{RR}}-f_{_{RQ}}(\rho_m-p_m)]
\\\label{38}&&-3a^2\phi_{_{t}}[2f_{_{RT}}-f_{_{TQ}}(\rho_m-p_m)]-3a^2\psi_{_{t}}
[2f_{_{RQ}}-f_{_{QQ}}(\rho_m-p_m)]=\mathcal{B}_{_{a}},
\\\nonumber&&\alpha(a^2f_{_{Q}}(\rho_m''-p_m'')-2(f_{_{R}}
-af_{_{Q}}(\rho_m'-p_m')-f_{_{Q}}\rho_m))+a\beta(-2f_{_{RR}}
+f_{_{RQ}}\\\nonumber&&\times(2\rho_m+a\rho_m'-ap_m'))+a\phi(-2f_{_{RT}}
+(2\rho_m+a\rho_m'-ap_m')f_{_{TQ}})+a\psi(-2f_{_{RQ}}\\\nonumber&&
+(2\rho_m+a\rho_m'-ap_m')f_{_{QQ}})+2a\alpha_{_{a}}((2\rho_m+a\rho_m'-ap_m')
f_{_{Q}}-2f_{_{R}})-a^2\beta_{_{a}}\\\nonumber&&\times(2f_{_{RR}}
-(\rho_m-p_m')f_{_{RQ}})-a^2\phi_{_{a}}(2f_{_{RT}}
-(\rho_m-p_m')f_{_{TQ}})-a^2\psi_{_{a}}(2f_{_{RQ}}\\\label{40}&&
-(\rho_m-p_m)f_{_{QQ}})+a\chi_{_{t}}(2f_{_{R}}
-3af_{_{Q}}(\rho_m'-p_m')-2f_{_{Q}}\rho_m)=0,\\\nonumber&&
\alpha(-4f_{_{RR}}+af_{_{RQ}}(\rho_m'-p_m')+2(\rho_m-p_m)f_{_{RQ}})
-a\beta(2f_{_{RRR}}-f_{_{RRQ}}(\rho_m-p_m))\\\nonumber&&
-a\phi(2f_{_{RRT}}-(\rho_m-p_m)f_{_{RTQ}})
-a\psi(2f_{_{RRQ}}-(\rho_m-p_m)f_{_{RQQ}})-a\alpha_{_{a}}
(2f_{_{RR}}\\\nonumber&&-f_{_{RQ}}(\rho_m-p_m))+2\alpha_{_{R}}((2\rho_m+a\rho_m'-ap_m')
f_{_{Q}}-2f_{_{R}})-a\beta_{_{R}}(2f_{_{RR}}
-f_{_{RQ}}\\\nonumber&&\times(\rho_m-p_m))-a\phi_{_{R}}(2f_{_{RT}}
-(\rho_m-p_m)f_{_{TQ}})-a\psi_{_{R}}(2f_{_{RQ}}
-(\rho_m-p_m)f_{_{QQ}})\\\label{41}&&+a\chi_{_{t}}(2f_{_{RR}}
-2f_{_{RQ}}(\rho_m-p_m)-f_{_{RQ}})=0,\\\nonumber&&
\alpha(-4f_{_{RT}}+af_{_{TQ}}(\rho_m'-p_m')+2(\rho_m-p_m)f_{_{TQ}})
-a\beta(2f_{_{RRT}}-f_{_{RTQ}}(\rho_m-p_m))\\\nonumber&&
-a\phi(2f_{_{RTT}}-(\rho_m-p_m)f_{_{TTQ}})
-a\psi(2f_{_{RTQ}}-(\rho_m-p_m)f_{_{TQQ}})-a\alpha_{_{a}}
(2f_{_{RT}}\\\nonumber&&-f_{_{TQ}}(\rho_m-p_m))+2\alpha_{_{T}}((2\rho_m+a\rho_m'-ap_m')
f_{_{Q}}-2f_{_{R}})-a\beta_{_{T}}(2f_{_{RR}}
-f_{_{RQ}}\\\nonumber&&\times(\rho_m-p_m))-a\phi_{_{T}}(2f_{_{RT}}
-(\rho_m-p_m)f_{_{TQ}})-a\psi_{_{T}}(2f_{_{RQ}}
-(\rho_m-p_m)f_{_{QQ}})\\\label{42}&&+a\chi_{_{t}}(2f_{_{RT}}
-2f_{_{TQ}}(\rho_m-p_m)-f_{_{TQ}})=0,\\\nonumber&&
\alpha(-4f_{_{RQ}}+af_{_{QQ}}(\rho_m'-p_m')+2(\rho_m-p_m)f_{_{QQ}})
-a\beta(2f_{_{RRQ}}-f_{_{RQQ}}(\rho_m-p_m))\\\nonumber&&
-a\phi(2f_{_{RTQ}}-(\rho_m-p_m)f_{_{TQQ}})
-a\psi(2f_{_{RQQ}}-(\rho_m-p_m)f_{_{QQQ}})-a\alpha_{_{a}}
(2f_{_{RQ}}\\\nonumber&&-f_{_{QQ}}(\rho_m-p_m))+2\alpha_{_{Q}}((2\rho_m+a\rho_m'-ap_m')
f_{_{Q}}-2f_{_{R}})-a\beta_{_{Q}}(2f_{_{RR}}
-f_{_{RQ}}\\\nonumber&&\times(\rho_m-p_m))-a\phi_{_{Q}}(2f_{_{RT}}
-(\rho_m-p_m)f_{_{TQ}})-a\psi_{_{Q}}(2f_{_{RQ}}
-(\rho_m-p_m)f_{_{QQ}})\\\label{43}&&+a\chi_{_{t}}(2f_{_{RQ}}
-2f_{_{QQ}}(\rho_m-p_m)-f_{_{QQ}})=0,\\\nonumber&&
\alpha(3a^2\{f-Rf_{_{R}}-Tf_{_{T}}-Qf_{_{Q}}-f_{_{T}}(\rho_m-3p_m)-\rho_m\}
-a^3\{f_{_{T}}(\rho_m'\\\nonumber&&-3p_m')-\rho_m'\})+a^3\beta(-Rf_{_{RR}}-Tf_{_{RT}}
-Qf_{_{RQ}}-(\rho_m-3p_m)f_{_{RT}})+a^3\phi\\\nonumber&&\times(-Rf_{_{RT}}-Tf_{_{TT}}
-Qf_{_{TQ}}-(\rho_m-3p_m)f_{_{TT}})+a^3\psi(-Rf_{_{RQ}}-Tf_{_{TQ}}\\\nonumber&&
-Qf_{_{QQ}}-(\rho_m-3p_m)f_{_{TQ}})+a^3\chi_{_{T}}(f-Rf_{_{R}}
-Tf_{_{T}}-Qf_{_{Q}}-f_{_{T}}(\rho_m\\\label{39}&&-3p_m)-\rho_m)=\mathcal{B}_{_{t}}.
\end{eqnarray}

In order to study the effect of extra force on strong curvature
regimes in cosmos, we consider different interactions between
curvature and matter variables. The non-geodesic equation of motion
reduces into standard geodesic equation for $T=0$, $Q=0$,
$f_T(R,T)=0$ and $f_{_{RT}}(R,T)=0$ with limit $p_m=0$. These
constraints lead to define $f(R,Q)$ and $f(R,T)$ models that
appreciate non-minimal interactions of Ricci scalar with contracted
Ricci, energy-momentum tensors and trace of energy-momentum tensor,
respectively. To study evolution of non-geodesic dust particles, we
consider $f_T(R,T)\neq0$ that yields $f(T,Q)$ model with $R=0$. It
would be interesting to investigate the behavior of extra force in
the presence of all three variables, i.e., $R, T$ and $Q$. Following
these conditions, we have
\begin{itemize}
\item $f(R,Q)$ Model, independent of $T$,
\item $f(R,T)$ Model, independent of $Q$,
\item $f(T,Q)$ Model, independent of $R$,
\item $f(R,T,Q)$ Model, with $R,~T,~Q\neq0$.
\end{itemize}
For the above possibilities of generic function, we solve non-linear
system of equations to construct symmetry generators, respective
conserved entities and exact solutions in the presence of dust fluid
(pressureless perfect fluid with $p_m=0$).

\section{The $f(R,Q)$ Model}

In this case, we consider that the generic function is independent
of trace $T$ and non-minimal coupling exists between curvature and
matter invariants $R$ and $Q$. For this assumption, we investigate
possible Noether point symmetries that lead to corresponding
conservation laws and exact solutions in the presence of boundary
term. We also construct cosmological analysis through some standard
cosmological measures and examine their behavior graphically. First,
we solve Eqs.(\ref{20})-(\ref{21}) and (\ref{22})-(\ref{37}) for
non-zero derivatives of generic function $f$ with $T=0$ and
$\mathcal{B}\neq0$, we obtain
$\alpha,_{_R},~\alpha,_{_T},~\alpha,_{_Q},~\chi,_{_a},
~\chi,_{_R},~\chi,_{_T},~\chi,_{_Q}=0$. For these constraints,
Eqs.(\ref{26})-(\ref{28}), (\ref{38}), (\ref{40}) and (\ref{42})
lead to the following form of generic function and unknown
coefficients of symmetry generators
\begin{eqnarray}\nonumber
&&\chi=c_6,\quad\alpha=c_7,\quad\beta=Z_1(t)Z_2(a)Z_3(R)Z_4(T)Z_5(Q),
\quad\phi=0,\quad\mathcal{B}=c_8,\\\nonumber
&&\psi=Z_1(t)Z_2(a)Z_3(R)Z_4(T)Z_5(Q)\frac{c_4Q+c_5}{c_4(R-c_2)},
\quad\rho_m(a)=Z_2(a),\\\label{44}
&&f(R,Q)=(-1)^{2c_1}c_3(R-c_2)^{c_1}(c_4Q+c_5)^{1-c_1}(c_1-1)^{c_1-1}.
\end{eqnarray}
Using above values of symmetry generator coefficients, density of
dust fluid and $f(R,Q)$ model in Eqs.(\ref{39}), (\ref{41}) and
(\ref{43}), we get
\begin{eqnarray*}\nonumber
Z_1(t)&=&c_9,\quad Z_2(a)=\frac{c_9}{a},\quad
Z_3(R)=c_{11}(R-c_2),\quad Z_4(T)=c_{10},\\\nonumber
Z_5(Q)&=&c_{12}(c_4Q+c_5)^{c_1},
\end{eqnarray*}
where $c_{i}~(i=1,...,12)$ denote arbitrary constants. Inserting
these solutions into Eq.(\ref{18}), the vector field becomes
\begin{eqnarray*}
Y=c_6\partial_{_t}+c_7\partial_{_{a}}+\frac{c_9^2c_{10}c_{11}c_{12}}{a}
(R-c_2)(c_4Q+c_5)^{c_1}\partial_{_{R}}+\frac{c_9^2c_{10}c_{11}c_{12}}{c_4}
(c_4Q+c_5)^{c_1+1}\partial_{_{Q}}.
\end{eqnarray*}
For the above solution of non-linear system of equations
(\ref{20})-(\ref{39}), the set of symmetry generators and respective
conserved quantities are
\begin{eqnarray}\nonumber
Y_1&=&\partial_{_t},\quad I_1=
a^3(f-Rf_{_{R}}-Qf_{_{Q}})-3f_{_{Q}}(2c_9a^2-\dot{a}^2)-
6a\dot{a}f_{_{R}}+3a\dot{a}f_{_{RQ}}\\\nonumber&\times&(c_9\dot{R}-2a\dot{Q})+
3a\dot{a}\dot{Q}(c_9f_{_{QQ}}-2af_{_{RQ}}),\\\nonumber
Y_2&=&\partial_{_{a}},\quad
I_2=12a\dot{a}f_{_{R}}+6a^2(\dot{R}f_{_{RR}}+\dot{Q}f_{_{RQ}}),
\\\nonumber Y_3&=&\frac{(R-c_2)(c_4Q+c_5)^{c_1}}{a}
\partial_{_{R}},\quad I_3=3a\dot{a}(R-c_2)(c_4Q+c_5)^{c_1}
(2af_{_{RR}}-c_9f_{_{RQ}}),\\\label{45}
Y_4&=&(c_4Q+c_5)^{c_1+1}\partial_{_{Q}},\quad
I_4=3a\dot{a}(c_4Q+c_5)^{c_1+1}(2af_{_{RQ}}-c_9f_{_{QQ}}).
\end{eqnarray}
It is interesting to mention here that the symmetry generators $Y_1$
and $Y_2$ ensure the presence of temporal translational and scaling
symmetries, respectively whereas $I_1$ refers to energy conservation
while $I_2$ identifies conservation of linear momentum. Solving
Eq.(\ref{45}) for $I_4=0$ with determined $f(R,Q)$ model (\ref{44}),
we obtain
\begin{equation}\label{ab}
a(t)=\frac{3^\frac{1}{4}}{\sqrt{c_2}e^{\sqrt{\frac{c_2}{3}}t}}
\sqrt{\sqrt{c_2}e^{\sqrt{\frac{c_2}{3}}t}(c_{14}-c_{13}
e^{2t\sqrt{\frac{c_2}{3}}})}.
\end{equation}
\begin{figure}\center{\epsfig{file=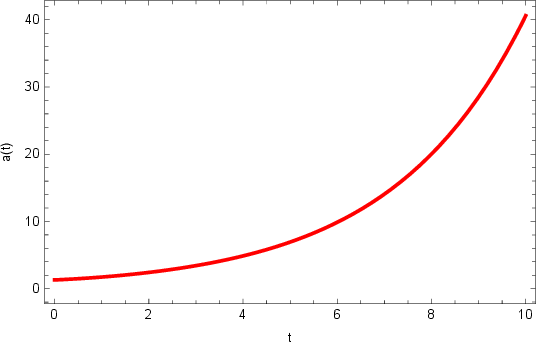,
width=0.43\linewidth}\epsfig{file=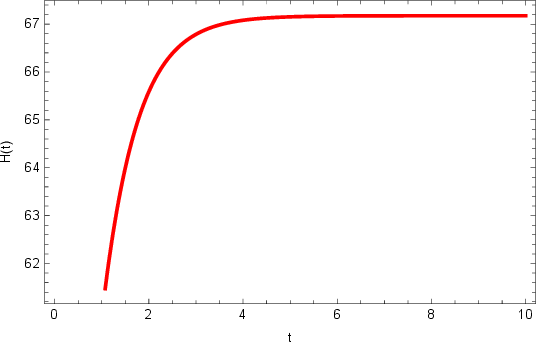,
width=0.43\linewidth}\\
\epsfig{file=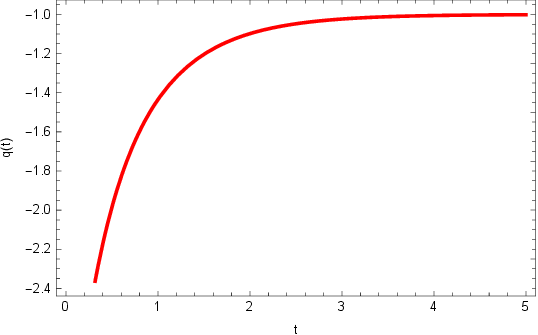,width=0.43\linewidth}
\epsfig{file=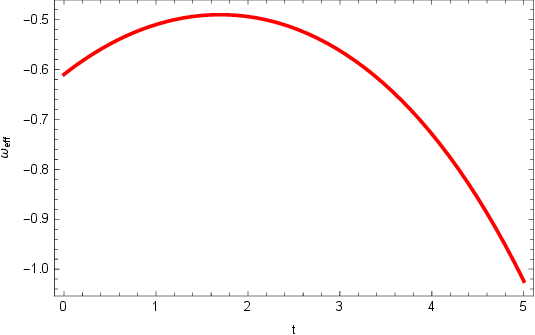,width=0.43\linewidth}}\caption{Plots of scale
factor, Hubble, deceleration and effective EoS parameters versus
cosmic time $t$ for $c_1=2$, $c_2=1.5$, $c_3=-1$, $c_4=0.5$,
$c_{5}=-1.2$, $c_{9}=3.9$, $c_{13}=-0.99$ and $c_{14}=0.2$.}
\end{figure}

To study cosmological impact of this exact solution, we establish
graphical analysis in Figure \textbf{1} (upper left plot) which
indicates that positively increasing trajectory corresponds
expanding cosmic state. To examine the state of expanding cosmos, we
discuss some important cosmological parameters, i.e., Hubble ($H$),
deceleration ($q$) and effective equation of state ($w_{eff}$)
parameters graphically. The Hubble parameter measures expansion rate
of cosmos while deceleration parameter ($q$) specifies the expansion
rate as accelerating ($q<0$), decelerating ($q>0$) or constant
($q=0$). The effective EoS parameter categorizes the accelerating
and decelerating cosmos into different regimes like radiation
($\omega=\frac{1}{3}$), matter ($\omega=0$), quintessence DE
($-1<\omega\leq-1/3$) and phantom DE ($\omega<-1$) dominated
regimes. The mathematical form of these standard parameters is given
as
\begin{equation}\nonumber
H=\frac{\dot{a}}{a},\quad q=-\frac{\dot{H}}{H^2}-1,\quad
w_{eff}=\frac{p_{_{eff}}}{\rho_{_{eff}}}=\frac{p_m+p_d}{\rho_m+\rho_d}.
\end{equation}
In the upper right plot of Figure \textbf{1}, the positively
evolving curve measures increasing rate of expansion which is found
to be consistent with current value of Hubble parameter, i.e.,
$H_0=67.69 \pm0.68$. Figure \textbf{1} (lower left plot) specifies
accelerated expansion whereas lower right plot refers to
quintessence DE era.

In modified theories of gravity, the minimally coupled $f(R)$ theory
of gravity provides the best explanation about enigmatic behavior of
cosmos. The framework of $f(R)$ gravity introduces such spectacular
models that unfolds mysteries behind early as well as current
universe via positive and negative powers of higher derivative of
curvature terms \cite{b5}. Besides these useful incentives, this
ghost-free theory may also illustrate unfeasible behavior due to
negative curvature terms. This issue can be eliminated by imposing
few constraints on higher-order derivatives, i.e.,
$f_R>0,~f_{_{RR}}>0$ with $R>R_0$, where $R_0$ refers to current
scalar curvature \cite{b6}. In non-minimally coupled $f(R,T)$
theory, Dolgov and Kawasaki suggested an additional constraint such
as $1+f(R,T)_{T}>0$. For $f(R,T,Q)$ gravity, the Dolgov-Kawasaki
instability analysis leads to the following additional viability
conditions
\begin{equation}\label{a}
\frac{1+f_{_{T}}+1/2Rf_{_{Q}}}{f_{_{R}}-\mathcal{L}_mf_{_{Q}}}>0,
\quad 3f_{_{RR}}+(1/2T-T^{00}f_{_{RQ}})\geq0.
\end{equation}

The squared speed of sound comes up with simple criteria to discuss
stable/unstable nature of new models. For positive squared speed of
sound, the linear perturbations become stable due to an oscillatory
motion whereas negative squared speed of sound confirms the presence
of strong perturbations leading to unstable state. The $r-s$
parameters provide a unique way to explore features of new models as
they build a compatibility between constructed and standard
cosmological models for their distinct values. For instance, the
established model admits a compatibility with $\Lambda$CDM, CDM
models and Einstein universe for ($r,s$)=(1,0), (1,0) and
(-$\infty$,$\infty$), respectively. If $s>0$ and $r<1$, then model
corresponds to quintessence and phantom DE eras while the
compatibility with Chaplygin gas model can be observed for $s<0$ and
$r>1$. Mathematically, the squared speed of sound and $r-s$
parameters are defined as
\begin{equation*}
v_s^2=\frac{\dot{p_d}}{\dot{\rho_d}},\quad
r=q+2q^2-\frac{\dot{q}}{H},\quad s=\frac{r-1}{3(q-\frac{1}{2})}.
\end{equation*}

We establish graphical analysis to study nature of constructed
$f(R,Q)$ model (\ref{44}) via squared speed of sound, viability
constraints and $r-s$ parameters. In Figure \textbf{2}, the upper
left plot shows positive evolution of function $f(R,Q)$ whereas
positively increasing squared speed of sound identifies stable state
of $f(R,Q)$ model in the upper right plot. In the lower panel of
Figure \textbf{2}, the viable nature of non-minimally coupled
$f(R,Q)$ model is observed as all conditions are preserved. The left
plot of Figure \textbf{3} determines compatibility of non-minimally
coupled $f(R,Q)$ model with $\Lambda$CDM model as $(r,s)=(1,0)$.
\begin{figure}\epsfig{file=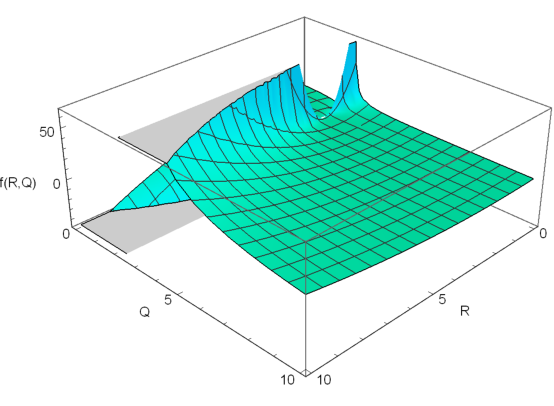,
width=0.44\linewidth}\epsfig{file=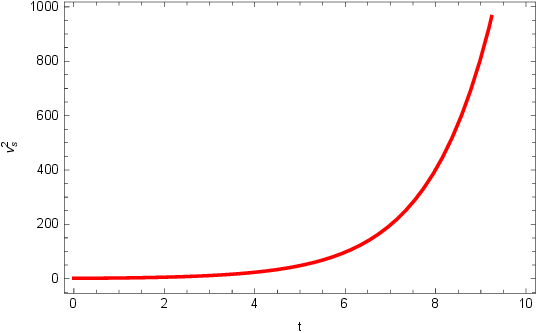,
width=0.44\linewidth}\\
\epsfig{file=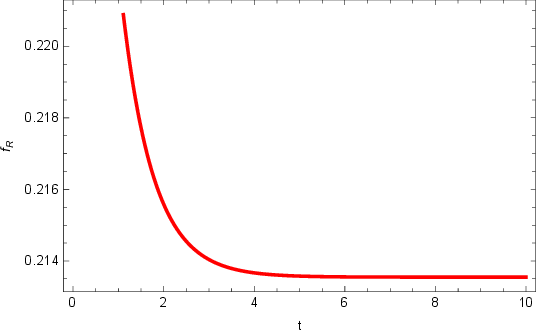,
width=0.44\linewidth}\epsfig{file=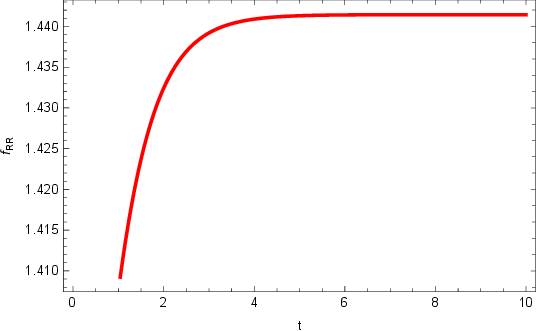,
width=0.44\linewidth}\\
\epsfig{file=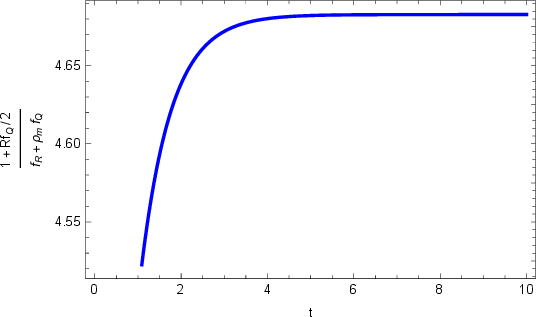,
width=0.44\linewidth}\epsfig{file=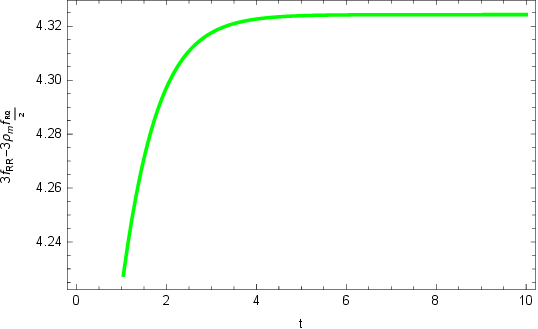,
width=0.44\linewidth}\caption{Plots of $f(R,Q)$ (upper left),
squared speed of sound (upper right) and viability conditions (lower
panel) versus cosmic time $t$.}
\end{figure}
\begin{figure}\epsfig{file=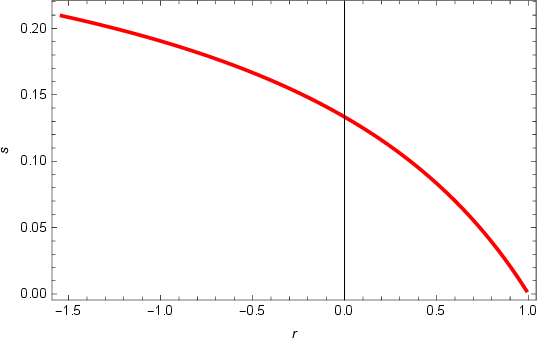,
width=0.44\linewidth}\epsfig{file=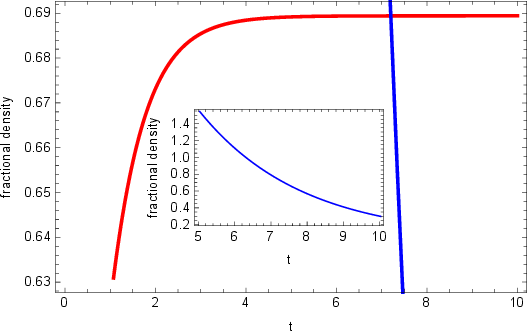,
width=0.44\linewidth}\caption{Plots of $r-s$ parameters (left) and
fractional densities (right) $\Omega_{m}$ (blue) and $\Omega_{d}$
(red) versus cosmic time $t$.}
\end{figure}

In the accelerate/decelerated expanding cosmos, the assessment of
fractional densities provides observational constraints to
counterbalance the contribution of ordinary matter and DE. For flat
cosmological model, the fractional densities are restricted to
follow $\Omega_m+\Omega_d=1$ where $\Omega_m\simeq0.3111$ and
$\Omega_d\simeq0.6889$ \cite{b8}. The mathematical formulation of
ordinary matter and DE fractional densities is given by
\begin{equation*}
\Omega_m=\frac{\rho_m}{3m_{pl}^2H^2},\quad
\Omega_d=\frac{\rho_d}{3m_{pl}^2H^2}.
\end{equation*}
The right plot of Figure \textbf{3} indicates that the fractional
densities relative to ordinary matter and DE preserve consistency
with Planck 2018 observational data as $\Omega_d=0.689$ and
$\Omega_m=0.303$ for $t=10$.

\section{The \textbf{$f(R,T)$ Model}}

Here, we discuss the impact of an indirect non-minimal coupling
between curvature and matter variables while the generic function
$f$ is assumed to be independent of the term $Q$ that induces a
direct non-minimal curvature-matter coupling. For this choice of
model, we solve the system of equations (\ref{20})-(\ref{43}) with
the restriction of non-zero derivatives of $f$ with $Q=0$ and
$\mathcal{B}\neq0$ that yield
\begin{eqnarray}\nonumber
\chi&=&Z_1(t),\quad\alpha=\frac{-Z_1(t)}{8c_1\sqrt{a}}+\frac{c_2}{\sqrt{a}},
\quad\beta=Z_3(R),\quad\phi=\frac{-Z_1(t)+3c_2}{3aZ_4(T),_{_{T}}},
\\\nonumber\mathcal{B}&=&Z_1(t)+a^{3/2}Z_1,_{_{t}},\quad\psi=0,\quad
f(R,T)=c_1R+Z_4(T),
\end{eqnarray}
where $c_1$ and $c_2$ refer to as constants while $Z_1,~Z_3$ and
$Z_4$ denote unknown functions of $t,~R$ and $T$, respectively. In
order to evaluate the solution of these unknown functions, we use
the above values of symmetry generator coefficients and $f(R,T)$
model in Eq.(\ref{39}) which gives
\begin{eqnarray*}\nonumber
Z_1(t)&=&c_3t+c_4,\quad Z_3(R)=c_9c_6,\quad
Z_4(T)=c_{7}T+c_8,\quad\rho_m(a)=\frac{c_8}{c_7}
-\frac{c_5a^{-3}}{c_7c_4}.
\end{eqnarray*}
Without loss of generality, we redefine $\frac{c_8}{c_7}=c_{10}$ and
$\frac{c_5}{c_4c_7}=c_{11}$. Inserting these solutions into
Eq.(\ref{18}), the respective vector field, boundary term and the
$f(R,T)$ model turn out to be
\begin{eqnarray}\nonumber
Y&=&(c_3t+c_4)\partial_{_t}+\left(-\frac{c_3t+c_4}{8c_1\sqrt{a}}
+\frac{c_2}{\sqrt{a}}\right)\partial_{_{a}}+c_9c_6\partial_{_{R}}
+\frac{3c_2-c_3t-c_4}{3ac_7}\partial_{_{T}},
\\\label{44a}\mathcal{B}&=&c_3t+c_4+c_3a^{3/2},\quad f(R,T)=c_1R+c_{7}T+c_8.
\end{eqnarray}
In this case, the $f(R,T)$ model admits an indirect curvature-matter
coupling due to linear and independent terms of Ricci scalar and
trace of the energy-momentum tensor. The Noether point symmetries
and associated first integrals are given by
\begin{eqnarray}\nonumber
Y_1&=&\frac{1}{\sqrt{a}}\partial_{_a}+\frac{1}{c_7a}\partial_{_T},\quad
I_1=6a^2\dot{a}f_{_{RR}},\\\label{45a} Y_2&=&\partial_{_{R}},\quad
I_2=6\sqrt{a}(2\dot{a}f_{_{R}}+a\dot{R}f_{_{RR}}+a\dot{T}f_{_{RT}})
+\frac{6a\dot{a}f_{_{RT}}}{c_7},
\\\nonumber Y_3&=&t\partial_{_t}-\frac{t}{8c_1\sqrt{a}}
\partial_{_{a}}-\frac{t}{3c_7a}\partial_{_{T}},
\quad I_3=1-a^3(f-Rf_{_{R}}-Tf_{_{T}}+\rho_m(1-f_{_{T}}))
\\\nonumber&-&6a\dot{a}^2f_{_{R}}
+\frac{12\sqrt{a}\dot{a}f_{_{R}}}{8c_1}\left(1-8c_1\sqrt{a}\dot{a}\right),
\\\nonumber Y_4&=&\partial_{_t}-\frac{1}{8c_1\sqrt{a}}
\partial_{_{a}}-\frac{1}{3c_7a}\partial_{_{T}},\quad
I_4=t+a^{3/2}-ta^3(f-Rf_{_{R}}-Tf_{_{T}}+\rho_m
\\\nonumber&\times&(1-f_{_{T}}))+6a\dot{a}^2tf_{_{R}}
-(12a\dot{a}f_{_{R}}+6a^2\dot{R}f_{_{RR}}+6a^2\dot{T}f_{_{RT}})\left(\dot{a}t
+\frac{t}{8c_1\sqrt{a}}\right)\\\nonumber&-&6a^2\dot{a}\dot{R}f_{_{RR}}
-6a^2\dot{a}tf_{_{RT}}(\dot{T}+\frac{1}{3ac_7}).
\end{eqnarray}
In the presence of boundary term, we obtain four symmetry generators
and corresponding Noether first integrals. These generators do not
appreciate temporal or scaling symmetries and consequently, the
respective conservation laws cannot be identified. For the
constructed $f(R,T)$ model (\ref{44a}), we solve Eq.(\ref{45a}) to
determine exact solution of the scale factor given by
\begin{equation*}
a(t)=\frac{1}{4c_1^2}\left[c_1^2(I_2t+8c_1c_{12})\right]^{\frac{2}{3}}.
\end{equation*}

Figure \textbf{4} interprets graphical analysis of cosmological
parameters for power-law scale factor. The upper left plot shows
positively increasing behavior of the scale factor specifying
expanding cosmos while the rate of expansion is found to be
decreasing as time passes leading to decelerated cosmic expansion
(upper right panel). In this case, the deceleration parameter also
corresponds to decelerated expanding universe as $q=1/2$. The
effective EoS parameter supports this analysis as
$0\leq\omega_{eff}<1$ (lower plot). Figure \textbf{5} (upper plots)
illustrates feasibility of constructed $f(R,T)$ model due to
positive squared speed of sound and viability conditions. The $r-s$
parameters fails to achieve compatibility of the established model
with any standard cosmological model as $(r,s)=(1,\infty)$. In the
lower plot, fractional density of DE dominates over matter density
while the dominance of matter energy density is observed as time
grows favoring decelerated expansion of the universe.
\begin{figure}\epsfig{file=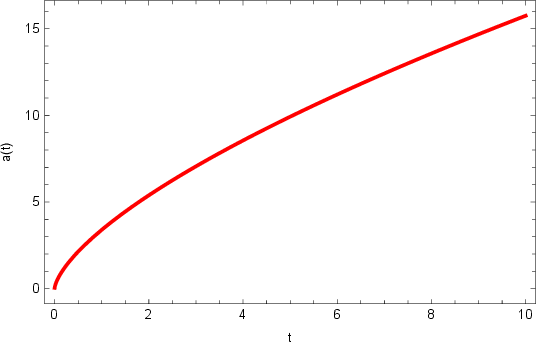,
width=0.39\linewidth}\epsfig{file=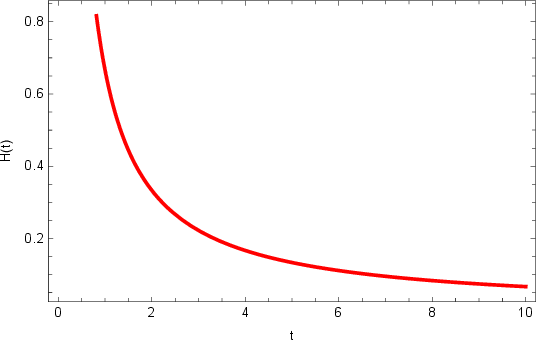,
width=0.39\linewidth}\\\center{\epsfig{file=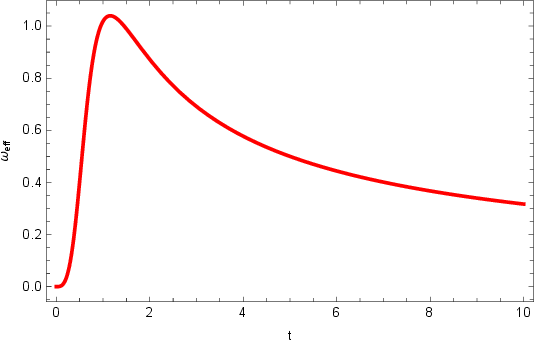,
width=0.39\linewidth}}\caption{Plots of scale factor, Hubble and
effective EoS parameters versus cosmic time $t$ for
$c_1=c_8=c_{11}=0.2$, $I_2=10$, $c_7=2$, $c_{10}=-1$ and
$c_{12}=0.01$.}
\end{figure}\begin{figure}\epsfig{file=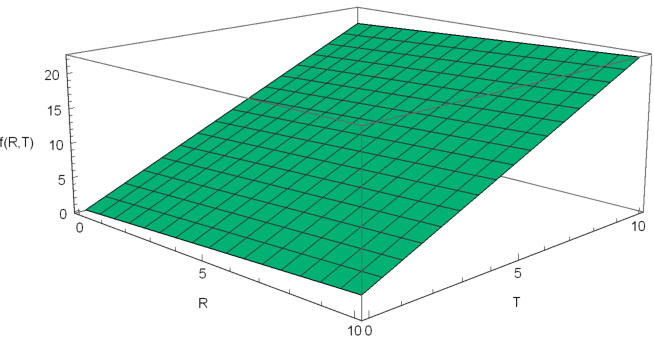,
width=0.39\linewidth}\epsfig{file=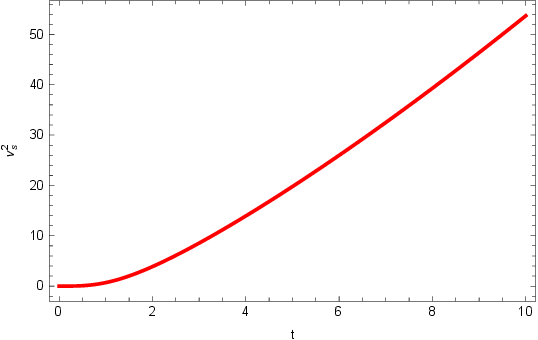,
width=0.39\linewidth}\\\center{\epsfig{file=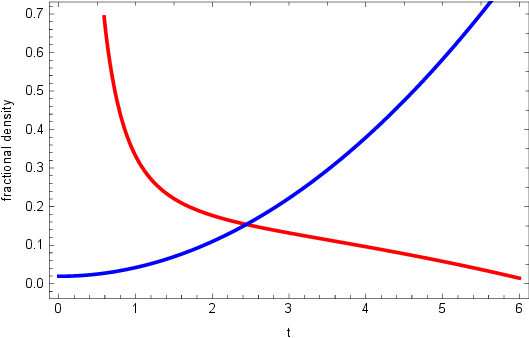,
width=0.39\linewidth}}\caption{Plots of $f(R,T)$ (left), squared
speed of sound (right) and fractional densities $\Omega_{m}$ (blue)
and $\Omega_{d}$ (red) versus cosmic time $t$.}
\end{figure}

\section{The \textbf{$f(T,Q)$ Model}}

In this section, we consider $T$ and $Q$ to be responsible for
strong non-minimal interactions in the absence of Ricci scalar. To
study the effect of such interaction on symmetries and conservation
laws, we choose $\beta=0$ and restrict generic function to admit
non-zero derivatives relative to $T$ and $Q$. The resulting solution
of over determining equations (\ref{20})-(\ref{43}) for
$\mathcal{B}\neq0$ is given as
\begin{eqnarray}\nonumber
&&\chi=c_7t+c_8,
\quad\phi=\frac{T-1/2}{a^2c_4}\left[a^2c_4(c_4\ln(\frac{2T-1}{2})+c_3)
-\frac{2c_5t}{c_3}+2c_4c_9\right],
\\\nonumber&&\mathcal{B}=c_5a+c_6,\quad\alpha=-\frac{c_7a}{3},
\quad\psi=c_{10}Q+c_{11},\quad\rho_m(a)=\frac{c_{12}\ln
a}{2},\\\label{44b}&&f(T,Q)=c_1T+c_3Q+c_4Q\ln(T-1/2)+c_2.
\end{eqnarray}
Here $c_i's~(i=1,2,...,12$) are constants whereas Eq.(\ref{44b})
indicates that the constructed $f(T,Q)$ model induces strong
non-minimal interactions. Inserting these solutions into
Eq.(\ref{18}) with $c_{13}=\frac{c_3c_7}{c_4}$, we obtain following
set of Noether point symmetries with conserved integrals given by
\begin{eqnarray}\nonumber
Y_1&=&\partial_{_t},\quad
I_1=-a^3(f-Tf_{_{T}}-Qf_{_{Q}}-\rho_m(1+f_T))+6a\dot{a}^2f_{_{Q}}
(2\rho_m+a\rho_m')\\\nonumber&+&3a^2\dot{a}\rho_m(\dot{T}f_{_{TQ}}
+\dot{Q}f_{_{QQ}}),\\\nonumber Y_2&=&Q\partial_{_{Q}},\quad
I_2=-3a^2\dot{a}Q\rho_mf_{_{QQ}},\quad Y_3=\partial_{_{Q}},\quad
I_3=-3a^2\dot{a}\rho_mf_{_{QQ}},
\\\nonumber Y_4&=&-\frac{2t(T-1/2)}{3c_4a^2}\partial_{_T},\quad
I_4=a+\frac{2t(T-1/2)\dot{a}\rho_mf_{_{TQ}}}{c_4}\\\nonumber
Y_5&=&\frac{2c_9(T-1/2)}{a^2}\partial_{_T},\quad
I_5=6(T-1/2)\dot{a}\rho_mf_{_{TQ}},\\\label{45b}
Y_6&=&(T-1/2)\partial_{_T},\quad
I_6=-3(T-1/2)a^2\dot{a}\rho_mf_{_{TQ}},\\\nonumber
Y_7&=&\partial_{_t}-\frac{a}{3}\partial_{_a}+(T-1/2)\ln(T-1/2)
\partial_{_T},\\\nonumber I_7&=&-a^3t(f-Tf_{_{T}}-Qf_{_{Q}}-\rho_m(1+f_T))
+a\dot{a}f_{_{Q}}(3\dot{a}t+2a)(2\rho_m+a\rho_m')\\\nonumber&+&
a^3\rho_m(\dot{T}f_{_{TQ}}+\dot{Q}f_{_{QQ}})-3a^2\dot{a}\rho_m
[(T-1/2)\ln(T-1/2)f_{_{TQ}}-t(\dot{T}f_{_{TQ}}\\\nonumber&+&\dot{Q}f_{_{QQ}})].
\end{eqnarray}
In the present case, we establish a set of seven symmetry generators
together with conserved entities. The Noether point symmetry
generator $Y_1$ ensures the presence of time translational symmetry
identifying energy conservation whereas $Y_2$ and $Y_6$ define
scaling symmetry relative to $Q$ and $T$ yielding conservation of
linear momentum. Furthermore, we formulate exact solution for the
scale factor by sorting out Eq.(\ref{45b}) that yields
\begin{equation*}
a(t)=c_{14}\exp\left[\frac{LambertW\left(-\frac{6I_6(t+c_{13}e^{-1})}
{c_{12}c_4}\right)}{3}+\frac{1}{3}\right]+c_{15}.
\end{equation*}
\begin{figure}
\center{\epsfig{file=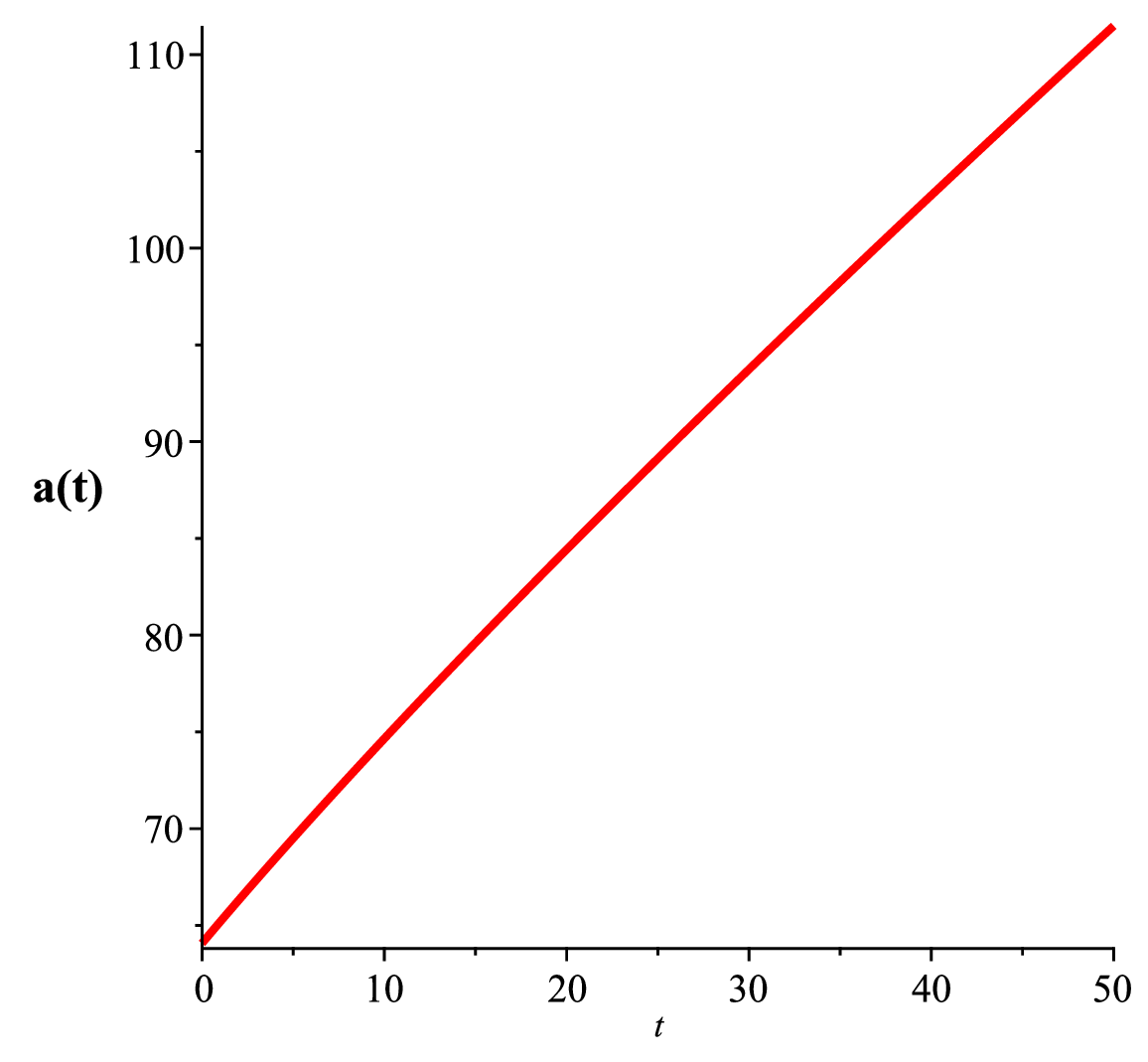,
width=0.27\linewidth}\epsfig{file=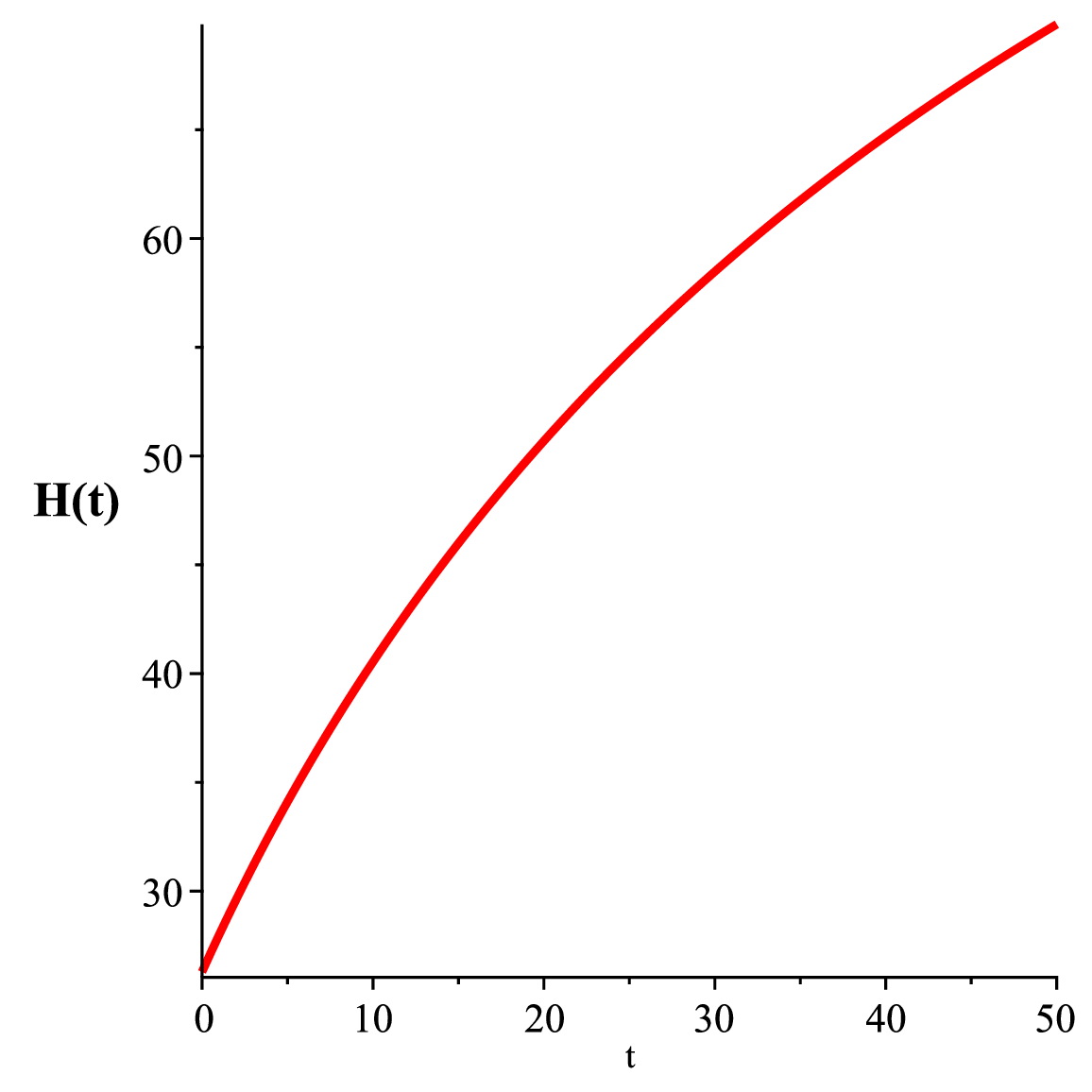,
width=0.27\linewidth}\\
\epsfig{file=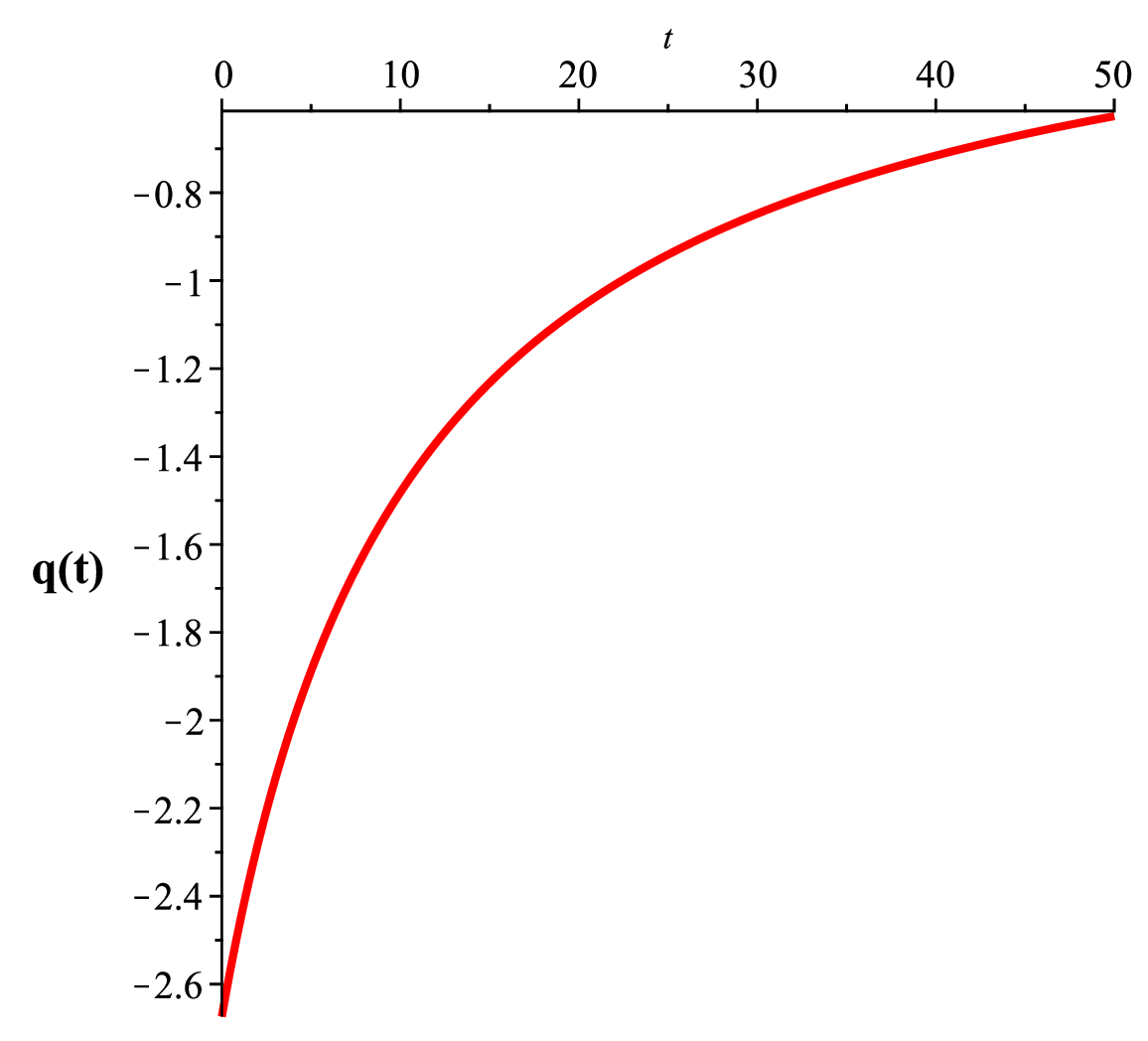,width=0.27\linewidth}
\epsfig{file=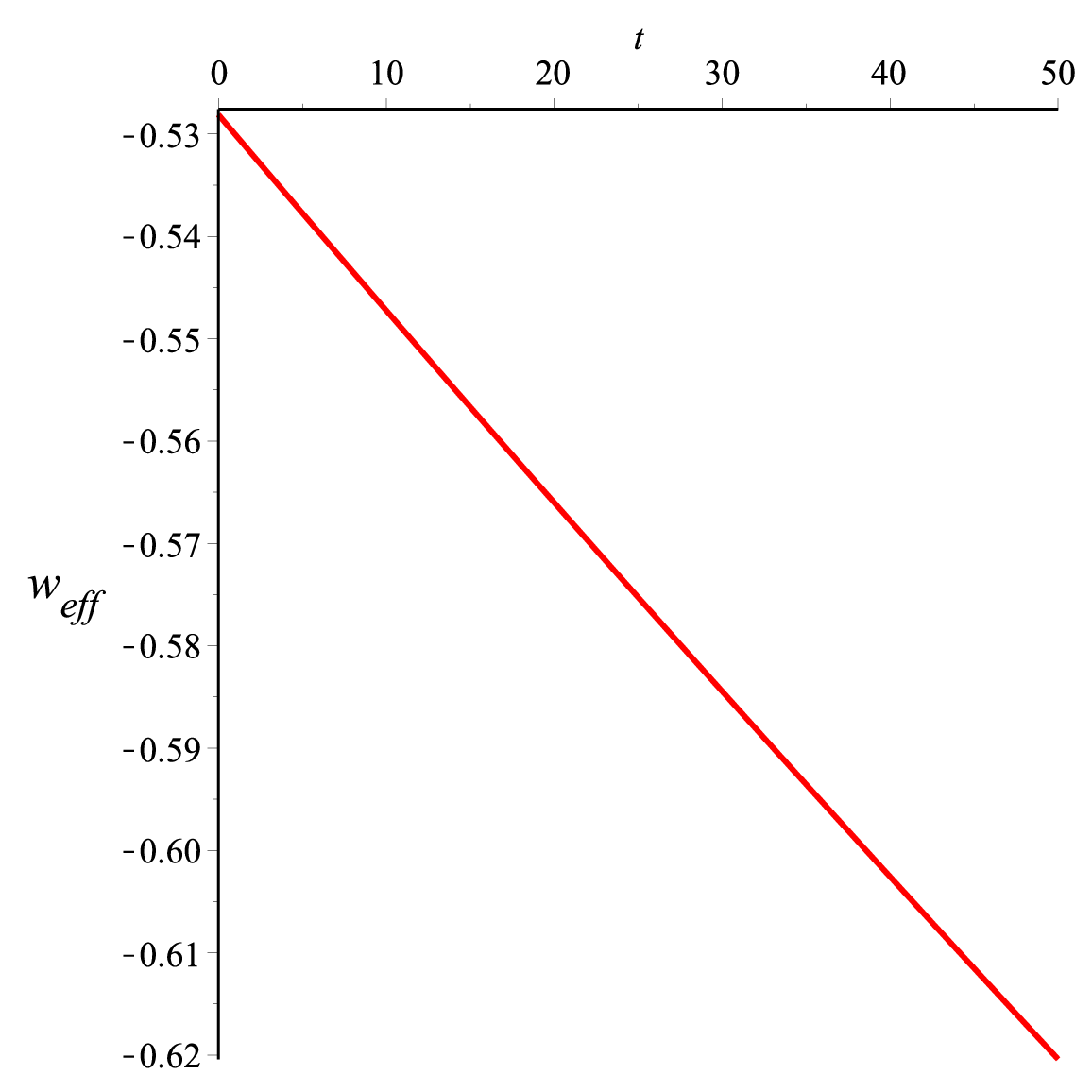,width=0.27\linewidth}}\caption{Plots of scale
factor, Hubble, deceleration and effective EoS parameters versus
cosmic time $t$ for $c_1=0.5$, $c_2=50$, $c_3=0.1$, $c_4=-0.18$,
$I_{6}=10$, $c_{12}=-1.2$, $c_{13}=10$, $c_{14}=0.1$ and
$c_{15}=50$.}
\end{figure}
\begin{figure}\center{\epsfig{file=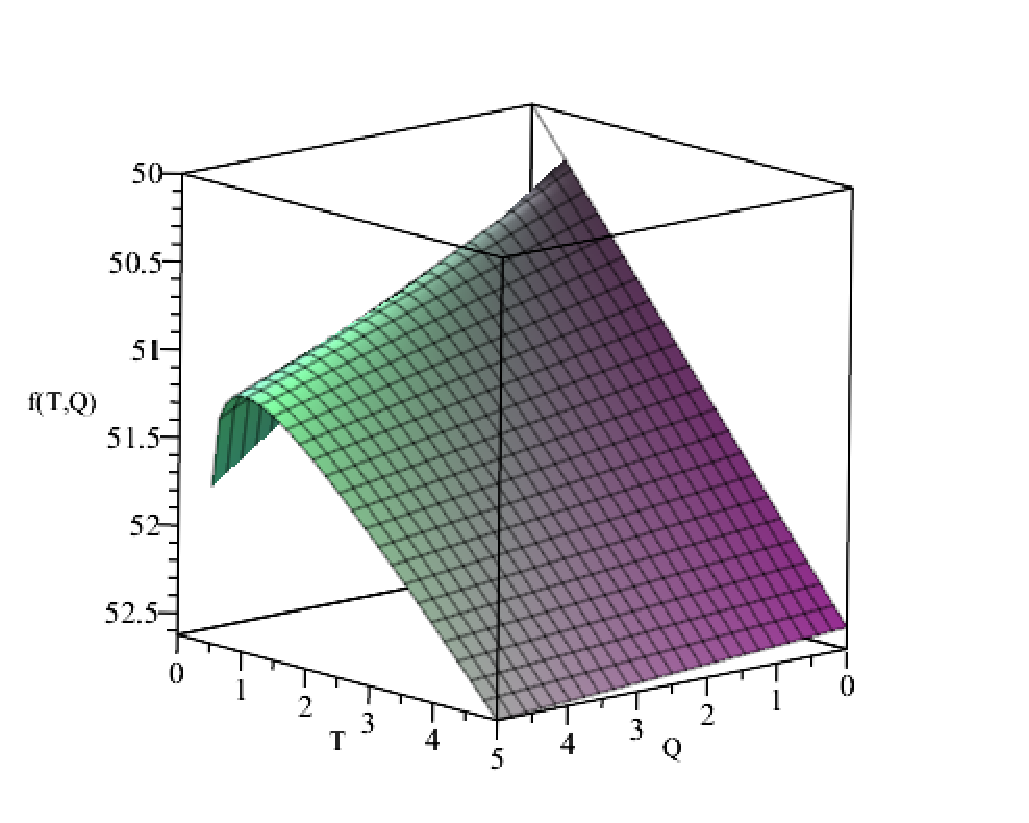,
width=0.5\linewidth}\epsfig{file=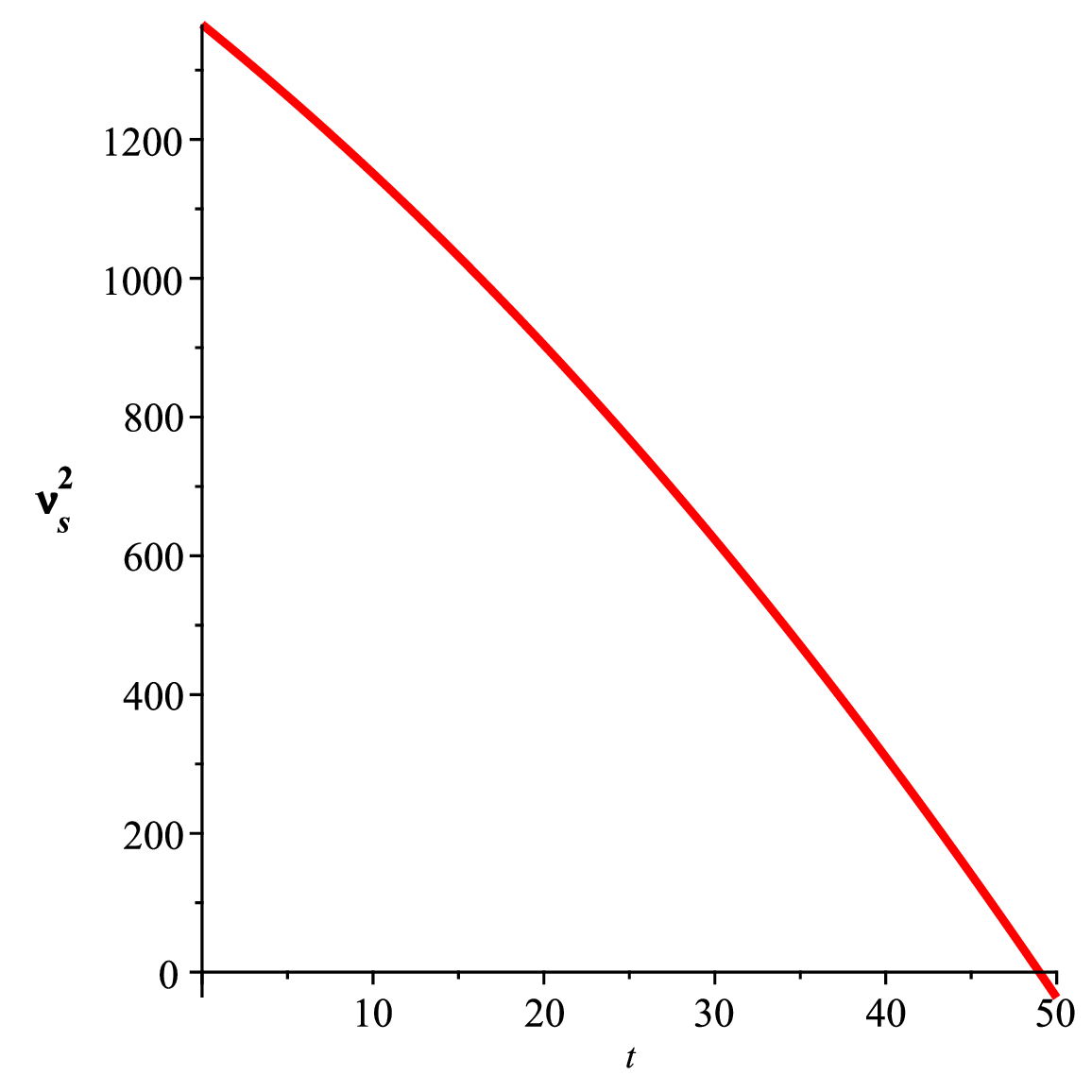,
width=0.35\linewidth}\\
\epsfig{file=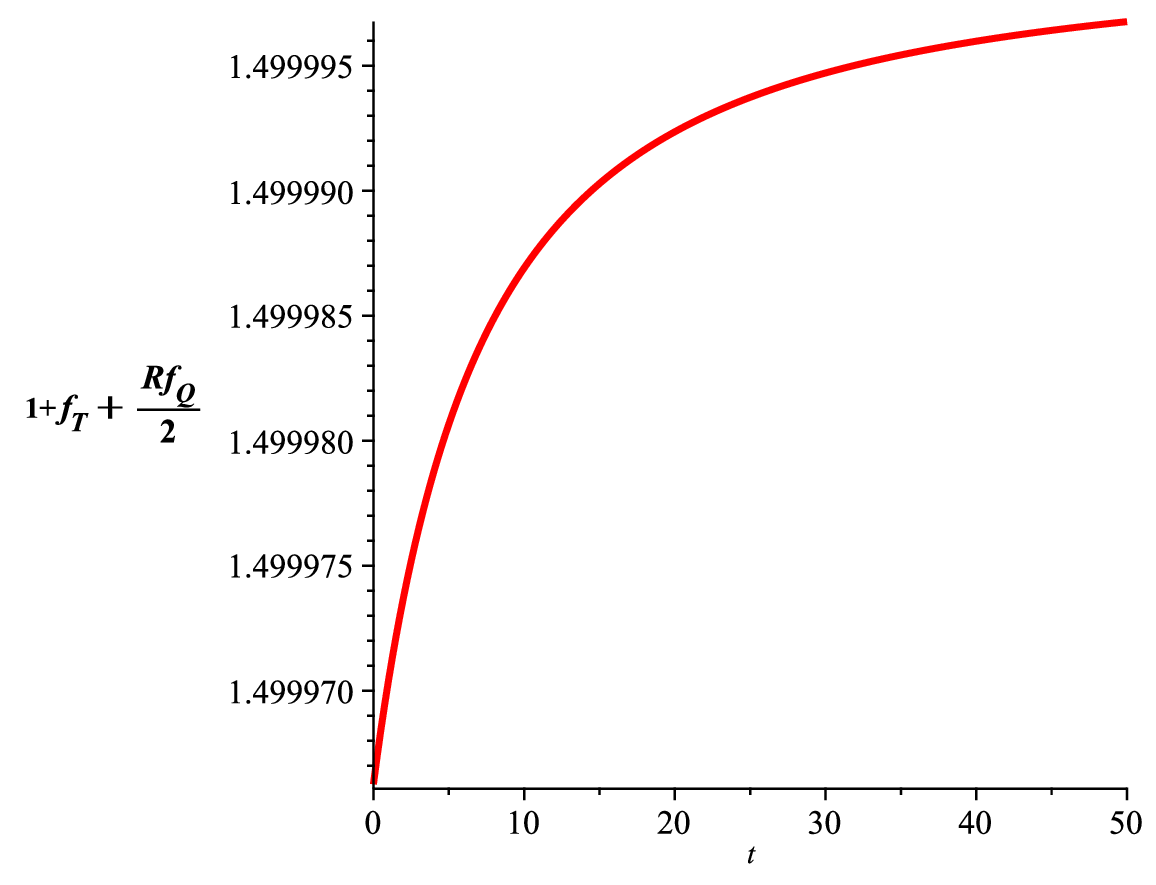,
width=0.4\linewidth}\epsfig{file=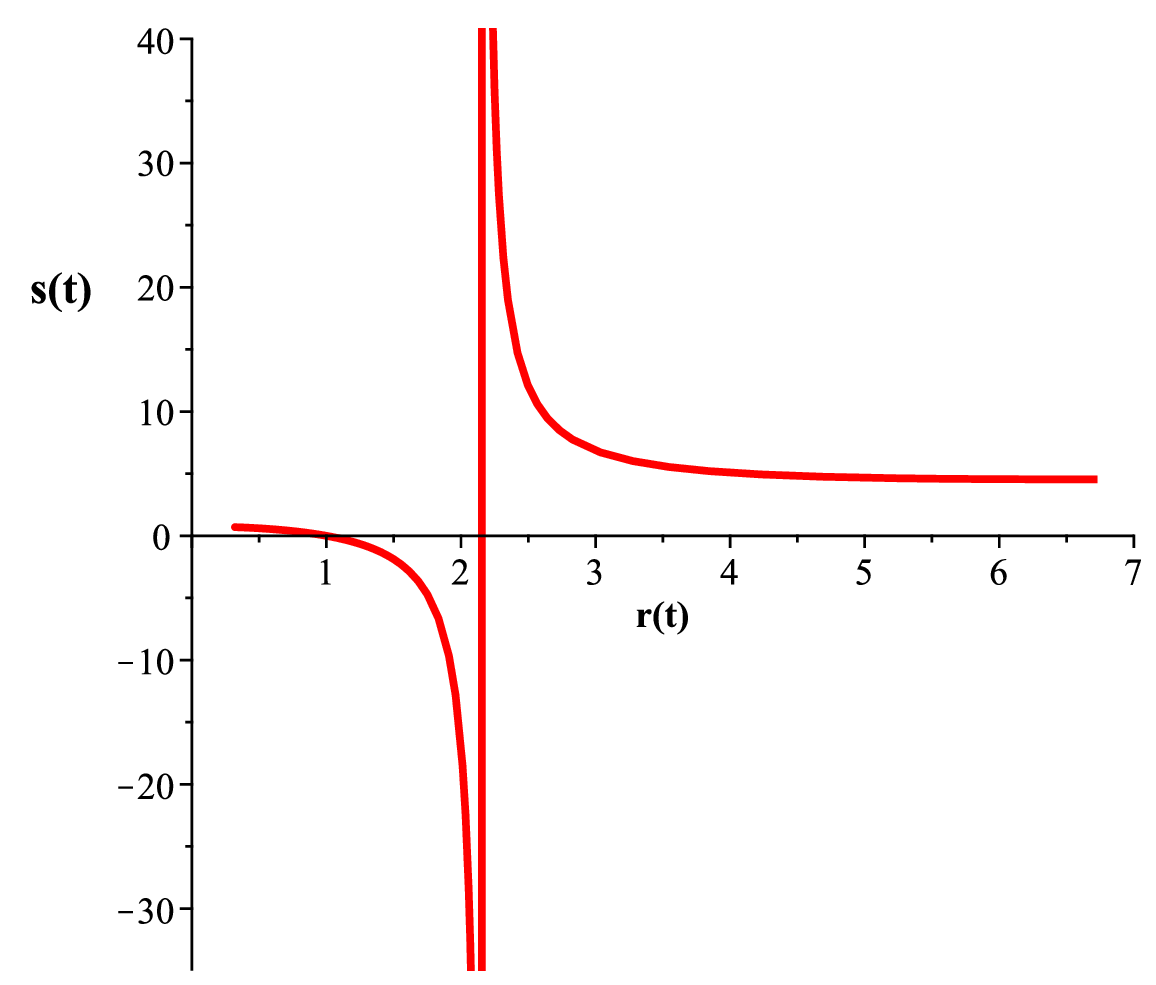,
width=0.35\linewidth}}\caption{Plots of $f(T,Q)$ (upper left),
squared speed of sound (upper right), viability condition (lower
left) and $r-s$ parameters (lower right) versus cosmic time $t$.}
\end{figure}
\begin{figure}\center{\epsfig{file=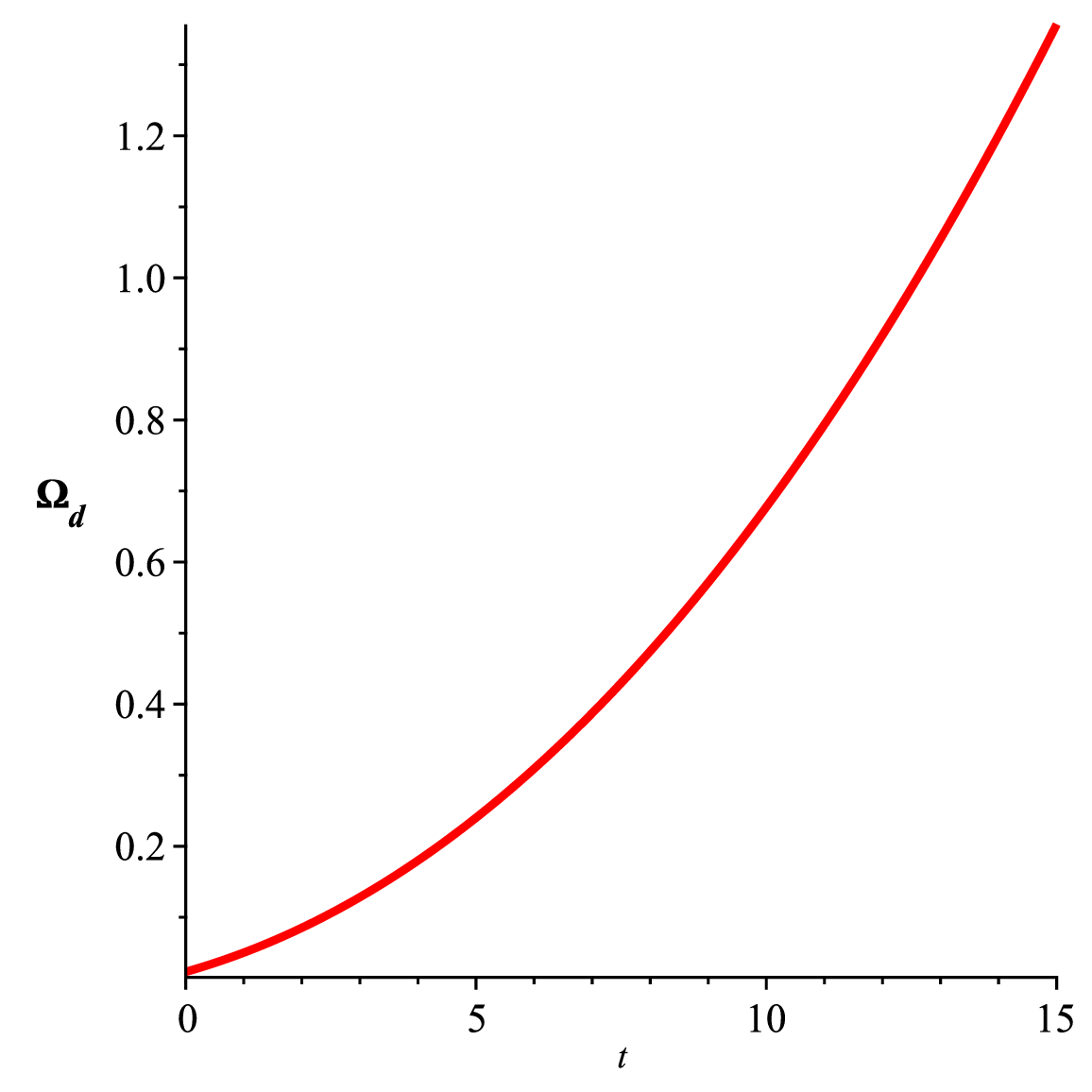,
width=0.4\linewidth}\epsfig{file=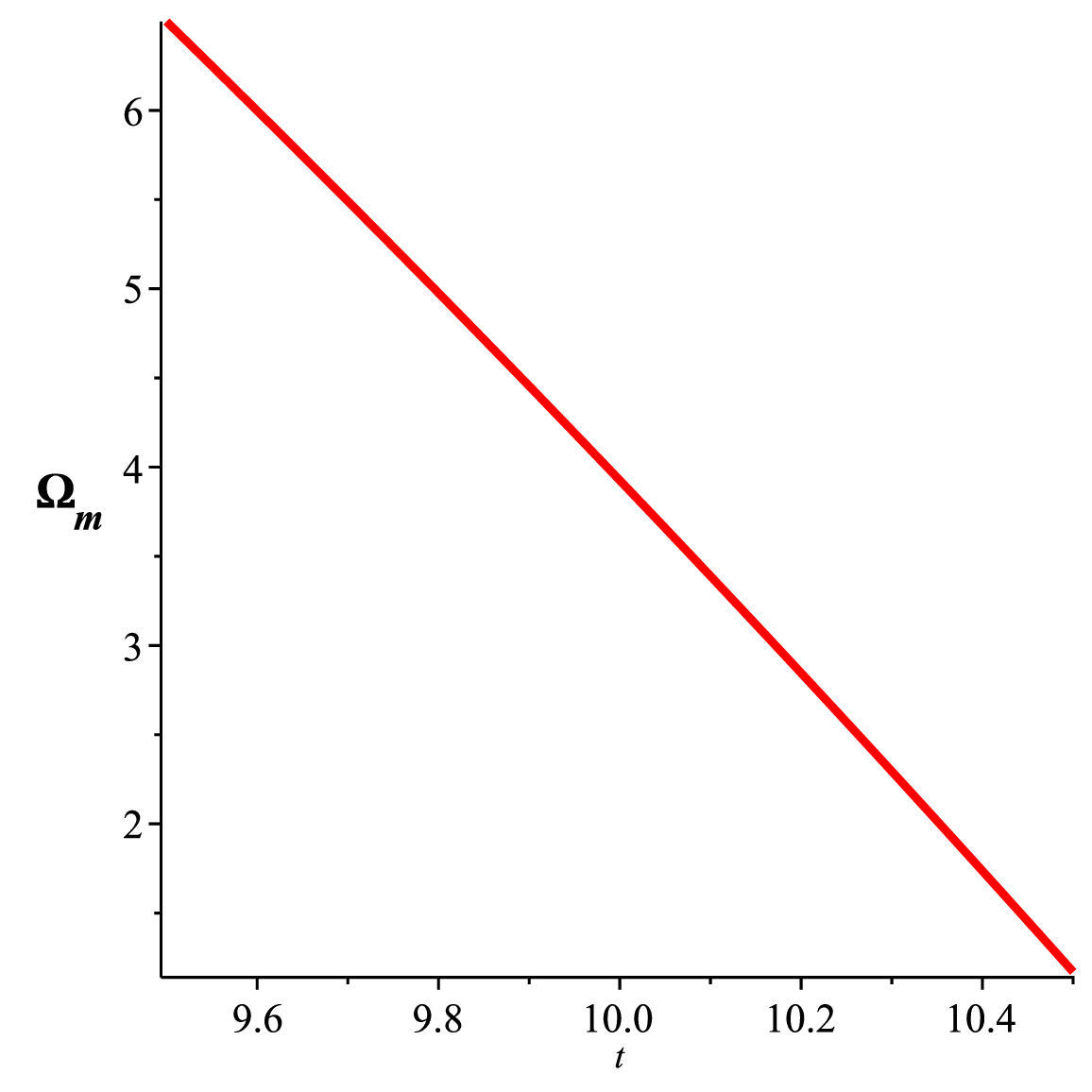,
width=0.4\linewidth}}\caption{Plots of fractional densities (right)
$\Omega_{m}$ (blue) and $\Omega_{d}$ (red) versus cosmic time $t$.}
\end{figure}

Figure \textbf{6} illustrates the evolution of scale factor (upper
left plot) while respective cosmological analysis is established via
Hubble (upper right), deceleration and effective EoS (lower panel)
parameters. The current value of Hubble parameter is achieved as
time grows and consequently, the graphical behavior of these
parameters supports accelerated quintessence DE era. Figure
\textbf{7} explores realistic nature of non-minimally coupled
$f(T,Q)$ model via squared speed of sound, viability condition and
$r-s$ parameters, respectively. In the given range of time, the
constructed model is found to be stable and viable whereas
initially, it preserves consistency with $\Lambda$CDM model. In
Figure \textbf{8}, the fractional density of DE is compatible with
Planck's constraints at $t=10$ whereas matter energy density exceeds
from the suggested value at the same time. Thus, the total
fractional energy density fails to achieve $\Omega_m+\Omega_d=1$.

\section{The \textbf{$f(R,T,Q)$ Model}}

In the present case, we determine Noether point symmetries with
conserved entities and also study the behavior of generic function
$f$ in the presence of non-minimally interacting scalar curvature
and matter variables. For non-zero boundary term, we consider
$\beta=0$, $\chi=\mathcal{B}=\alpha=\varphi(t,a)$ and restrict
generic function to have non-zero first order derivatives relative
to independent variables $R,~T$ and $Q$. Inserting these constraints
into Eqs.(\ref{20})-(\ref{38}), we obtain
\begin{eqnarray}\nonumber
&&\chi=F_1(t),\quad\mathcal{B}=F_2(t),\quad\alpha=F_3(a),
\quad\phi=F_4(a,R,T,Q),\\\nonumber&&f(R,T,Q)=F_5(T)R+QF_6(T)+F_(7).
\end{eqnarray}
Here $F_i's,~(i=1,2,...,7)$ are unknown functions. For above
solutions, we solve Eqs.(\ref{40})-(\ref{39}) which yield
\begin{eqnarray}\nonumber
&&F_1(t)=c_2,\quad F_3(a)=\frac{c_1}{\sqrt{a}},\quad F_5(T)=c_3,
\quad\rho_m=\frac{c_4}{a^{\frac{3}{2}}},
\\\nonumber&&\psi=-[3c_1a^{\frac{3}{2}}(QTF_{_{6T}}
+TF_{_{7T}}-F_{_{7}})+a^3QTF_4F_{_{6TT}}+a^3QF_4F_{_{6T}}
+a^3TF_4\\\nonumber&&F_{_{7TT}}+F_{_{2t}}]/[a^3TF_{_{6T}}].
\end{eqnarray}
Verifying the resulting solution for $F_6=F_7$, we get
\begin{equation*}
F_2(t)=c_5t,\quad F_6(T)=c_6T+c_7,\quad
F_4(a,R,T,Q)=\frac{9Tc_8+4c_9a}{4a\sqrt{a}}.
\end{equation*}
The explicit from of $f(R,T,Q)$ model turns out to be
\begin{equation}\label{44c}
f(R,T,Q)=c_3R+(Q+1)(c_6T+c_7).
\end{equation}
It is worth noting that the constructed model induces strong
non-minimal interactions between $T$ and $Q$ whereas the Noether
symmetries with boundary and associated conserved integral are given
by
\begin{eqnarray}\nonumber Y_1&=&\partial_{_t},\quad
I_1=-a^3(f-Rf_R-Tf_{_{T}}-Qf_{_{Q}}-\rho_m(1+f_T))\\\nonumber&-&6a\dot{a}^2f_{_{R}}
-3a^2\dot{a}\rho_m\dot{T}f_{_{TQ}}+9a^2\dot{a}^2\rho_m'f_{_{Q}}),\\\nonumber
Y_2&=&\frac{1}{\sqrt{a}}\partial_{_{a}}-\frac{3Q}{4a\sqrt{a}}
\partial_{_{Q}},\\\nonumber I_2&=&12\sqrt{a}\dot{a}f_{_{R}}
-6\sqrt{a}\dot{a}f_{_{Q}}(2\rho_m+a\rho_m')
+3a\sqrt{a}\dot{T}\rho_mf_{_{TQ}},\\\label{48c}
Y_3&=&\frac{9}{16a\sqrt{a}}\left(4T\partial_{_{T}}
-Q\partial_{_{Q}}\right),\quad
I_3=-\frac{27a^2\dot{a}\rho_mf_{_{TQ}}T}{4a\sqrt{a}},
\\\nonumber Y_4&=&\frac{1}{T\sqrt{a}}\left(T\partial_{_{T}}
-Q\partial_{_{Q}}\right),\quad
I_4=-\frac{3a^2\dot{a}\rho_mf_{_{TQ}}}{\sqrt{a}}.
\end{eqnarray}
Here, we formulate four symmetries together with Noether first
integrals while there is only one symmetry generator that
corresponds to time translational symmetry and respective Noether
integral specifies energy conservation. For constructed $f(R,T,Q)$
model (\ref{44c}), the conserved integral $I_3$ yields
\begin{equation*}
a(t)=\frac{(-9c_4^2c_6(-9c_{10}c_4^2c_6+2I_3t)^2)^{\frac{2}{3}}}
{(-9c_{10}c_4^2c_6+2I_3t)^2}.
\end{equation*}
\begin{figure}\center{\epsfig{file=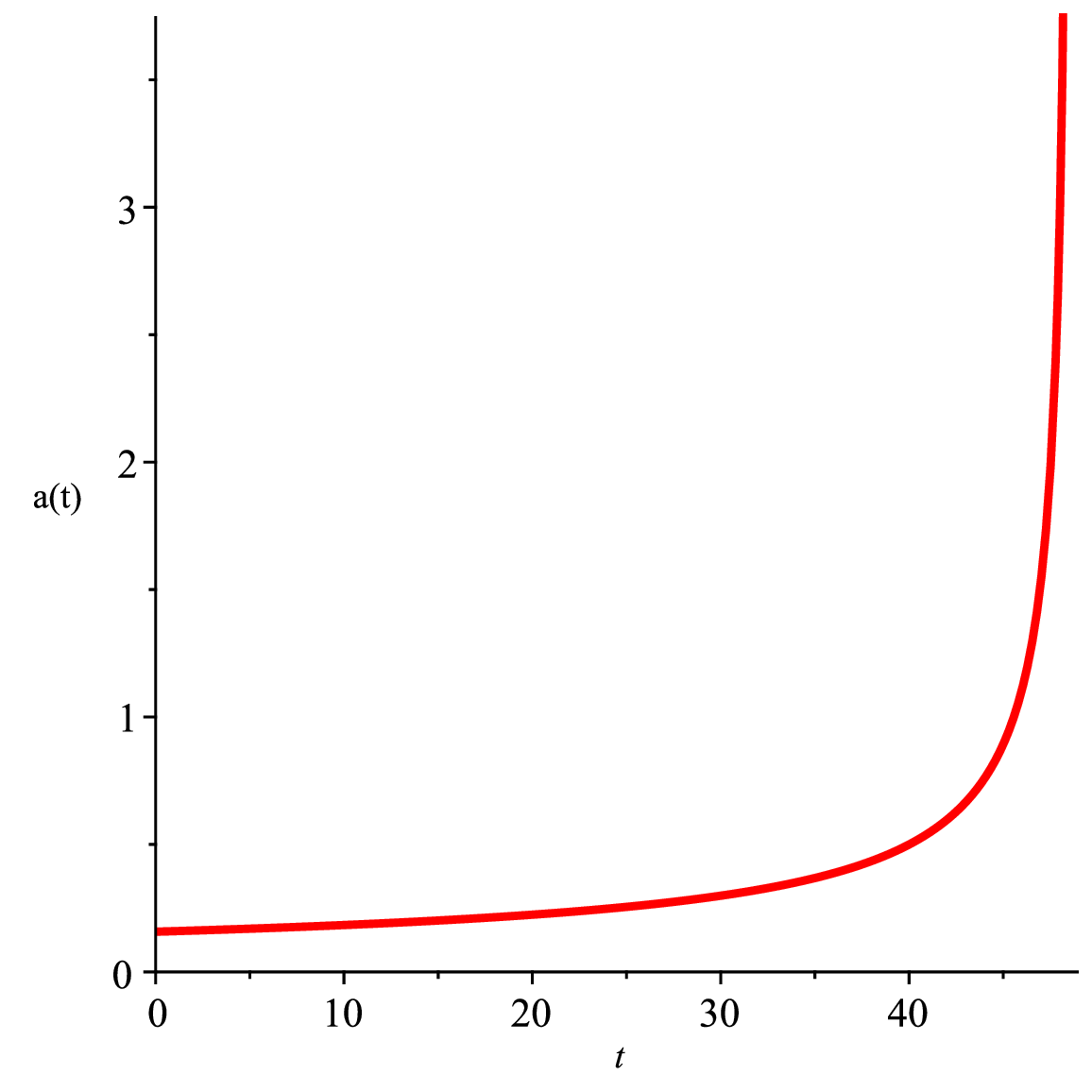,
width=0.27\linewidth}\epsfig{file=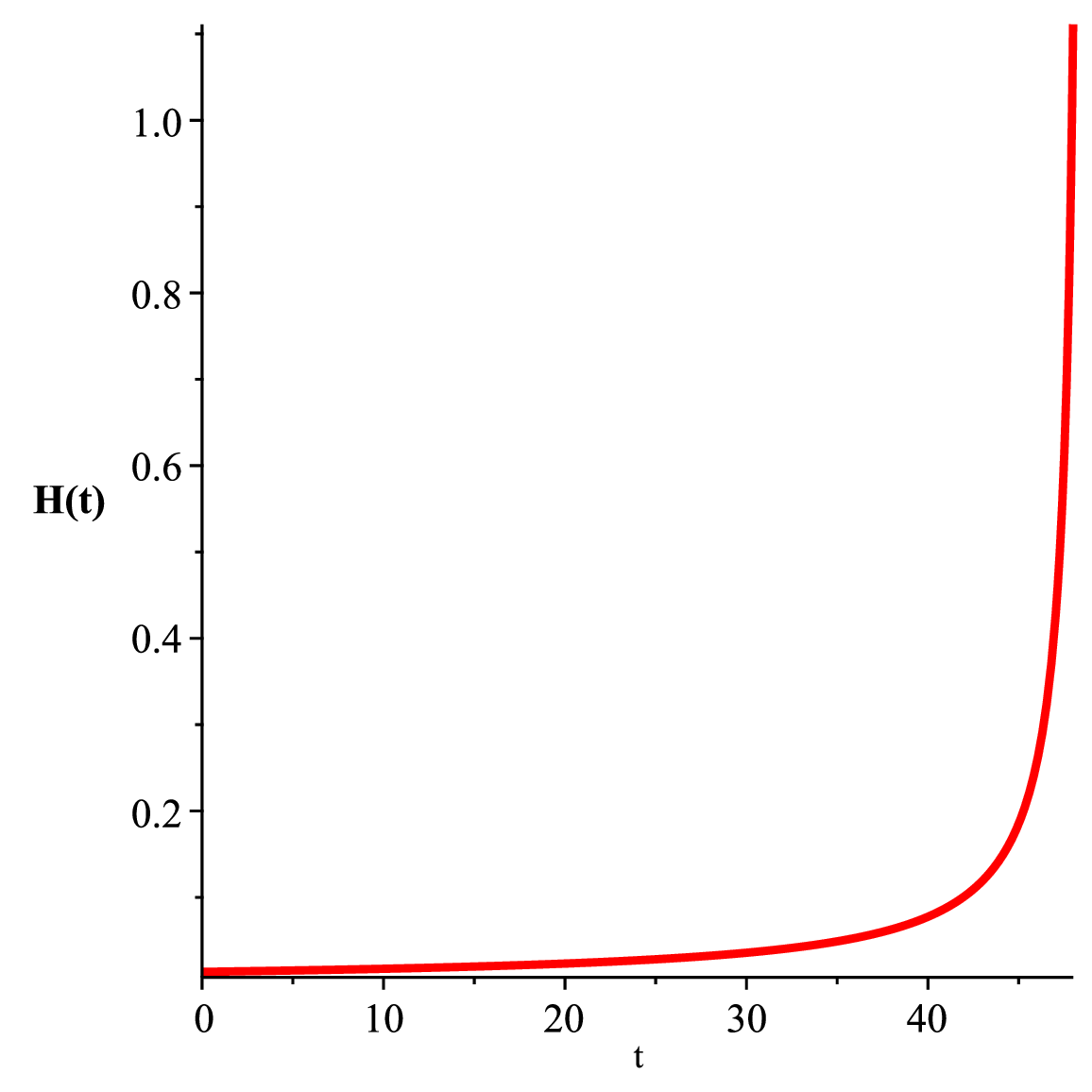,
width=0.27\linewidth}\\
\epsfig{file=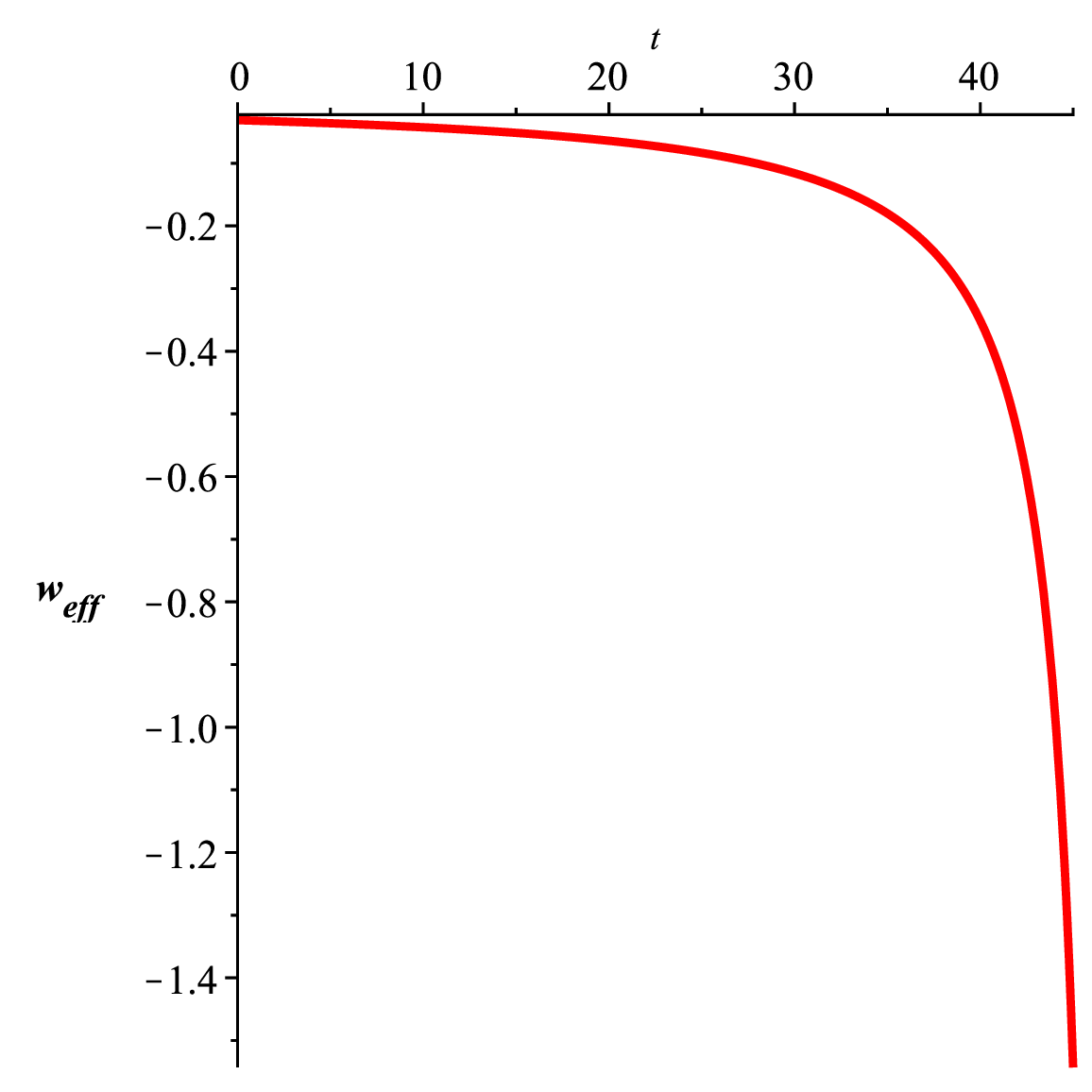,width=0.27\linewidth}}\caption{Plots of scale
factor, Hubble and effective EoS parameters versus cosmic time $t$
for $c_3=15$, $c_4=0.5$, $c_6=-13.5$, $c_7=1.5$, $I_{3}=5$ and
$c_{10}=-16$.}
\end{figure}
\begin{figure}\center{\epsfig{file=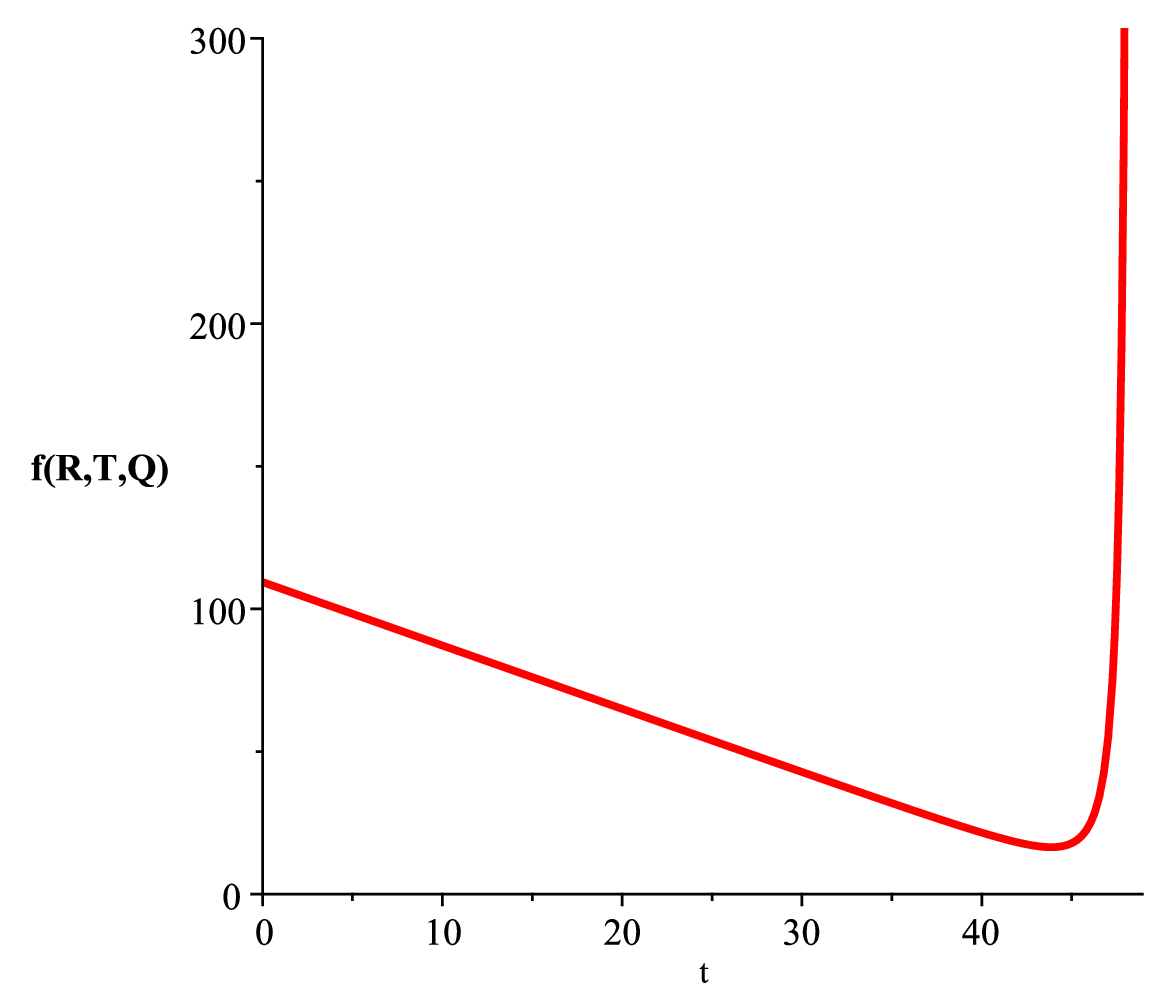,
width=0.3\linewidth}\epsfig{file=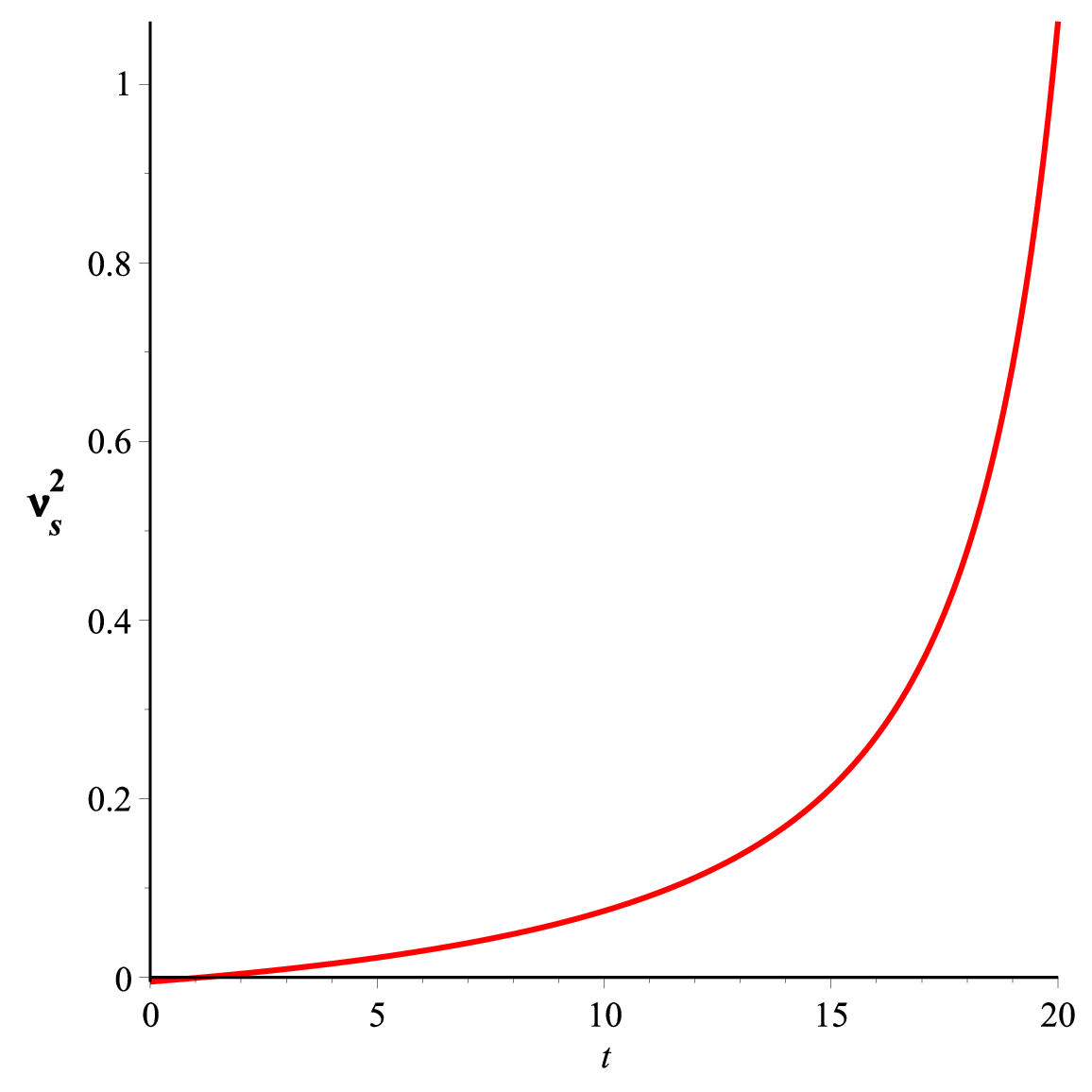,
width=0.3\linewidth}\\
\epsfig{file=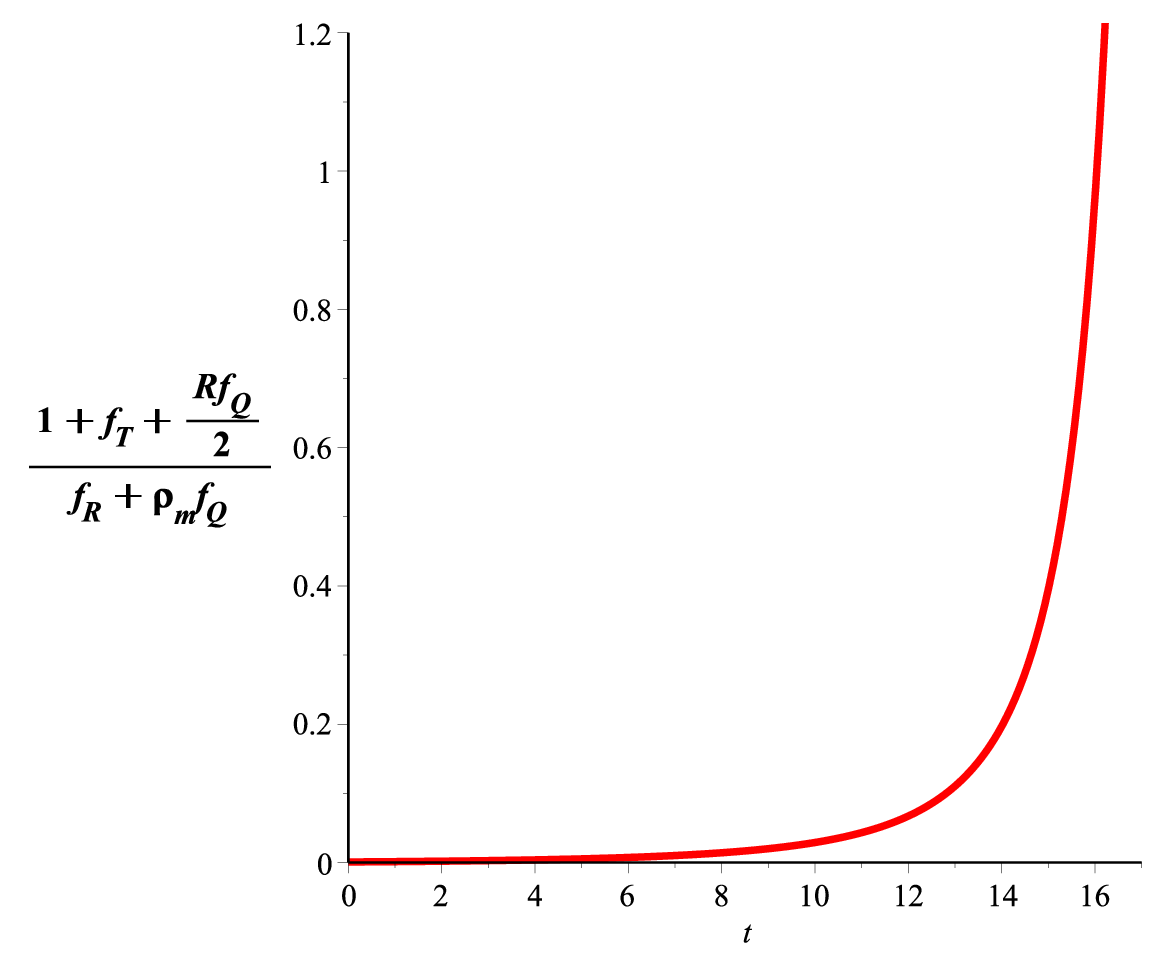, width=0.4\linewidth}}\caption{Plots of
$f(R,T,Q)$ (upper left), squared speed of sound (upper right) and
viability condition (lower plot) versus cosmic time $t$.}
\end{figure}
\begin{figure}\center{\epsfig{file=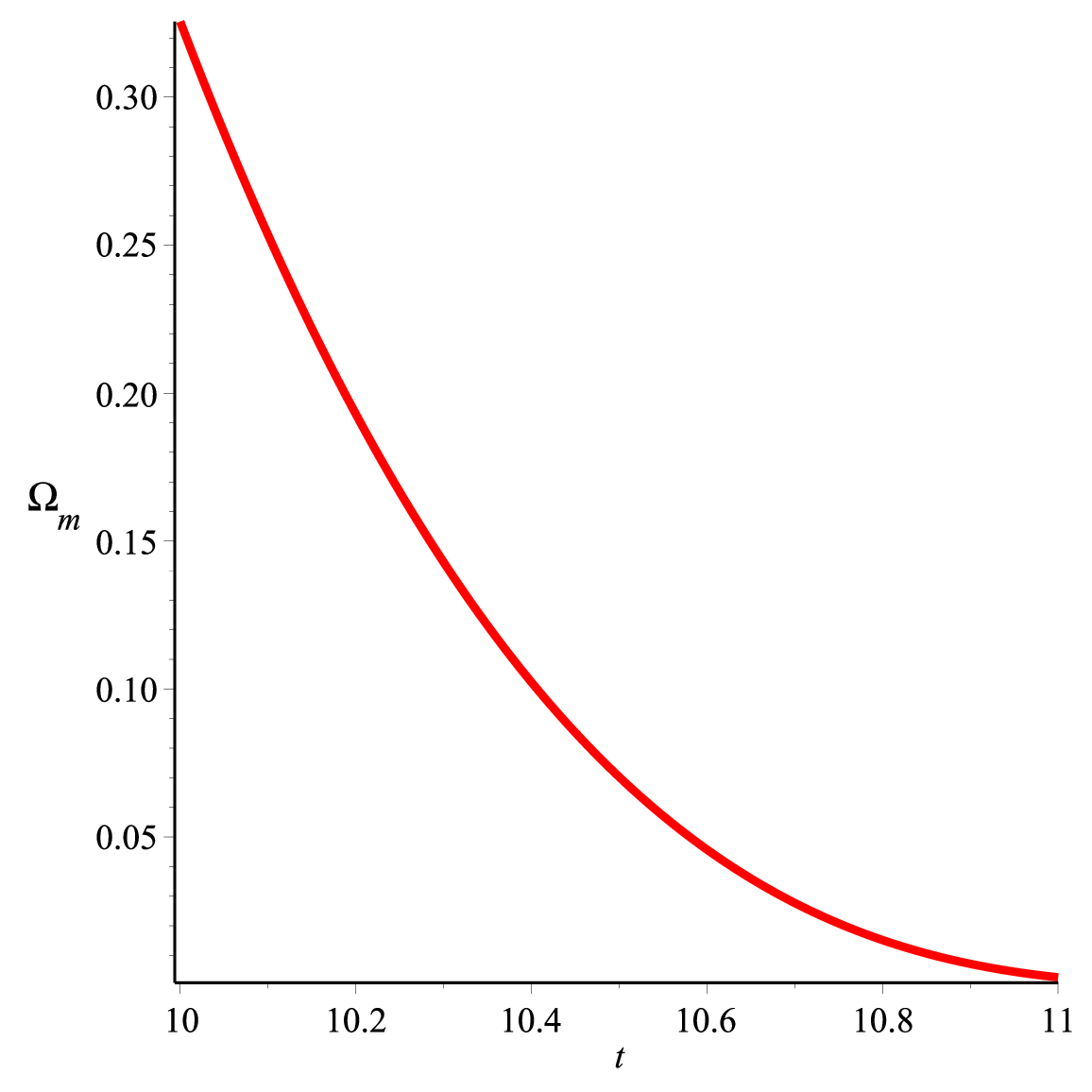,
width=0.35\linewidth}\epsfig{file=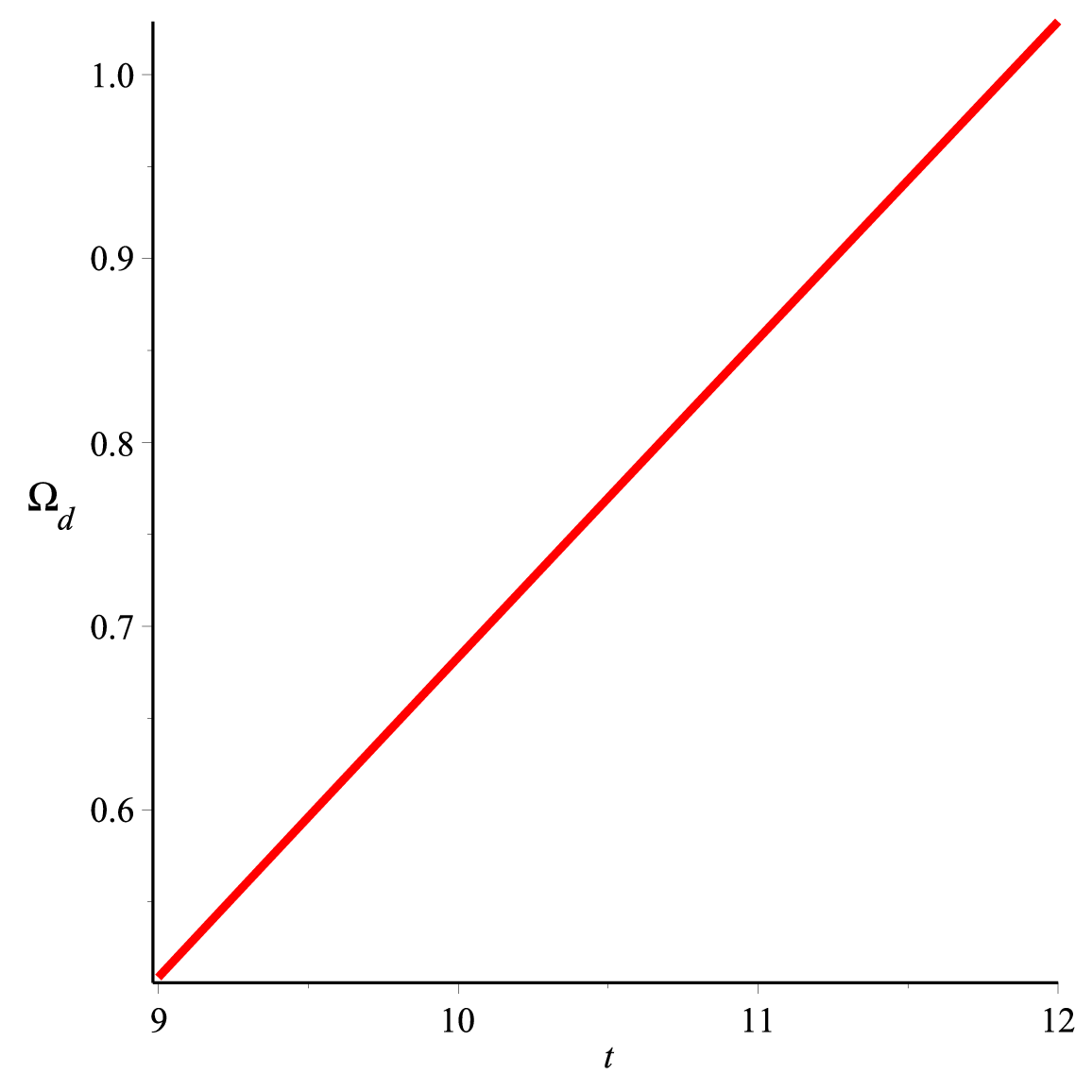,
width=0.35\linewidth}}\caption{Plots of fractional densities
$\Omega_{m}$ (left) and $\Omega_{d}$ (right) versus cosmic time
$t$.}
\end{figure}

In Figure \textbf{9}, we discuss cosmological impact of the above
exact solution through standard parameters which indicates that the
universe meets up with accelerated expansion. Furthermore, the
negative value of deceleration parameter ($q=-2.5$) ensures
accelerated expansion of the universe. The graphical analysis of
effective EoS parameter illustrates that the cosmos enters into
quintessence DE era and gradually joins phantom DE phase (lower
plot). For numerical value of deceleration parameter, the
$(r,s)=(3.75,0.38)$ parameters characterize Chaplygin gas model.
Figure \textbf{10} identifies stable as well viable behavior of
positively evolving non-minimally coupled $f(R,T,Q)$ model in the
background of DE. In Figure \textbf{11}, the fractional energy
densities follow observational constraints in the given time
interval.

\section{Final Remarks}

The revolutionary idea of non-minimal coupling between geometry and
matter sectors leads to fascinating approaches that explore
different cosmological scenarios as well as cosmic evolution from
its origin to the current state. The present work is devoted to
study evolutionary stages of flat and isotropic cosmos filled with
dust distribution in the background of $f(R,T,Q)$ gravity. For this
purpose, we have followed Noether Gauge symmetry technique that
induces temporal or spatial symmetries along with relevant
conservation laws, i.e., energy or linear/angular momentum
conservation, respectively. Besides symmetries and respective
conservation laws, this intriguing technique also helps to formulate
exact solution to understand corresponding cosmological impact.

The presence of strong or weak non-minimal interactions between
curvature and matter variables investigate different cosmological
phenomenologies. In the this work, we have considered different
choices for generic function $f(R,T,Q)$ admitting non-minimal
interactions such as $f(R,Q)$, $f(R,T)$ and $f(T,Q)$ models
independent of $T,~Q$ and $R$, respectively. For each choice and
general $f(R,T,Q)$ model, we have found explicit forms of generic
function as well as discussed the existence of Noether point
symmetries together with conservation laws in the presence of
boundary term. The summary for formulated symmetries and respective
conservation laws for $\mathcal{B}\neq0$
is given in Table \textbf{1}.\\
\textbf{Table 1:} Symmetries and Conservation laws for
$\mathcal{B}\neq0$.
\begin{table}[bht]
\centering
\begin{small}
\begin{tabular}{|c|c|c|c|}
\hline\textbf{Model}&\textbf{Symmetry}&\textbf{Conservation Laws}\\
\hline $f(R,Q)$&Temporal $\&$ scaling&Energy $\&$ Linear Momentum\\
\hline$f(T,Q)$&Temporal $\&$ scaling&Energy $\&$ Linear Momentum\\
\hline$f(R,T,Q)$&Temporal&Energy conservation\\
\hline
\end{tabular}
\end{small}
\end{table}\\
It is interesting to mention here that for $f(R,T)$ model with
$\mathcal{B}\neq0$, the system of over determining equations fail to
produce any well-known symmetry and conservation law. In each case,
the conserved integrals yield exact solutions for scale factor. We
have investigated cosmological features of these solutions via
graphical analysis of some standard cosmological parameters, i.e.,
Hubble, deceleration and effective EoS parameters. Planck 2018
suggested different values of $H$ at $68\%CL$ given by \cite{b8}
\begin{eqnarray*}\nonumber
H_0&=&67.27\pm0.60\quad\text{(TT+TE+EE+low E)},\\\nonumber
H_0&=&67.36\pm0.54\quad\text{(TT+TE+EE+low E+lensing)},\\\nonumber
H_0&=&67.66\pm0.42\quad\text{(TT+TE+EE+low E+lensing+BAO)}.
\end{eqnarray*}

Furthermore, we have examined stable/unstable and viable/unviable
state of constructed $f(R,Q),~f(R,T),~f(T,Q)$ and $f(R,T,Q)$ models
through squared speed of sound and viability conditions suggested by
Dolgov-Kawasaki instability analysis. The contribution of matter
content is also studied by analyzing fractional densities of dust
fluid and DE. The observational values of $\Omega_m$ and $\Omega_d$
with $68\%$ percent limit are given as \cite{b8}
\begin{eqnarray}\nonumber
\Omega_m&=&0.3166\pm0.0084\quad\text{(TT+TE+EE+low E)},\\\nonumber
\Omega_m&=&0.3158\pm0.0073\quad\text{(TT+TE+EE+low
E+lensing)},\\\nonumber
\Omega_m&=&0.3111\pm0.0056\quad\text{(TT+TE+EE+low E+lensing+BAO)}
\\\nonumber \Omega_d&=&0.7116\pm0.0084\quad\text{(TT+TE+EE+low
E)}, \\\nonumber \Omega_d&=&0.6847\pm0.0073\quad\text{(TT+TE+EE+low
E+lensing)}, \\\nonumber
\Omega_d&=&0.68889\pm0.0056\quad\text{(TT+TE+EE+low E+lensing+BAO)}
\end{eqnarray}
The consistency of all parameters is checked against Planck's 2018
observational data. The results are summarized as follows.

\begin{itemize}
\item \textbf{$f(R,Q)$ Model}
\end{itemize}
In the presence of boundary term, the graphical analysis of
cosmological parameters supports accelerated cosmic expansion as
current value of Hubble parameter is achieved, deceleration
parameter is negative while effective EoS parameter corresponds to
quintessence DE phase. The constructed non-minimally coupled
$f(R,Q)$ model is found to be stable, viable and compatible with
$\Lambda$CDM model. The fractional densities are consistent with
Planck suggested constraints. Zubair et al. \cite{r1} investigated
cosmic evolution using particular $f(R,Q)$ model and power-law
Hubble parameter without using Noether Symmetry approach. They
constructed model constraints that referred to $\Lambda$CDM limit
and explains current accelerated expansion.
\begin{itemize}
\item \textbf{$f(R,T)$ Model}
\end{itemize}
The cosmological parameters are found to be in favor of decelerated
expanding cosmos due to decreasing rate of expansion, positive
deceleration parameter and correspondence of effective EoS parameter
with radiation dominate era. The stable as as well as viable
$f(R,T)$ model admits minimal coupling between scalar curvature and
matter though it does not appreciate compatibility with any standard
cosmological model. In the background of radiation-dominated phase,
the fractional density relative to dust fluid dominates over DE
fractional density ensuring decelerated expanding cosmos. We have
discussed the presence of Noether Gauge symmetry in $f(R,T)$ theory
\cite{r2}. For this purpose, we considered minimally coupled
$f(R,T)$ model that admits symmetry generator relative to energy
conservation whereas exact solution for the scale factor is not
measured. In the present work, we have found $f(R,T)$ model
appreciating minimal interactions between curvature and matter
variables while power-law exact solution is obtained that defines
decelerated expanding cosmos.
\begin{itemize}
\item \textbf{$f(T,Q)$ Model}
\end{itemize}
The graphical study of cosmological parameters leads to accelerated
expansion of the universe for Noether Gauge symmetries. We have
formulated non-minimally coupled $f(T,Q)$ model that preserves
stability and viability conditions while consistency with
$\Lambda$CDM model is achieved for $\mathcal{B}\neq0$. The
fractional densities are inconsistent with accelerated expanding
cosmos for non-zero boundary term.
\begin{itemize}
\item \textbf{$f(R,T,Q)$ Model}
\end{itemize}
In this case, the graphical interpretation demonstrates accelerated
expansion with a transition from quintessence to phantom regions.
The established $f(R,T,Q)$ model incorporates non-minimal coupling
between $T$ and $Q$ while appreciates minimal coupling with $R$. The
model is found to be feasible as $v_s^2>0$ and viability condition
are satisfied. This realistic model admits compatibility with
Chaplygin gas model whereas graphical illustration of fractional
densities also correspond to accelerated expanding cosmos.

We have found symmetry generators and conserved quantities for each
case except for $f(R,T)$ model, where symmetries and respective
conservation laws do not correspond to any standard symmetry or
conserved entity. The constructed $f(R,T,Q)$ models are found to be
stable and viable. The compatibility of established models with
$\Lambda$CDM and Chaplygin gas models is observed. We conclude that
the constructed solutions favor accelerated cosmic expansion
whenever generic function involves non-minimal coupling with $Q$.\\\\
\textbf{Data Availability Statement:} No new data were created or
analyzed in this study.

\end{document}